\documentclass[journal]{IEEEtran}

\usepackage{float}
\usepackage{blindtext}
\usepackage{graphicx}
\usepackage{amssymb}
\usepackage[cmex10]{amsmath}
\usepackage{gensymb}
\usepackage{MnSymbol}
\usepackage{amsthm}
\usepackage{multirow}

\DeclareMathOperator*{\argmin}{arg\,min}

\usepackage{color}
\usepackage{comment}
\usepackage{siunitx}
\newtheorem{Definition}{Definition}

\usepackage{todonotes}   % notes showed

\newcommand*{\vect}[1]{\mathbf{#1}}
\newcommand*{\greekvect}[1]{\boldsymbol{#1}}
\usepackage{cite}
\ifCLASSINFOpdf
\else
\fi
\begin{document}
%
% paper title
% can use linebreaks \\ within to get better formatting as desired
\title{Identifiability and Parameter Estimation of the Single Particle Lithium-Ion Battery Model}
%
%
% author names and IEEE memberships
% note positions of commas and nonbreaking spaces ( ~ ) LaTeX will not break
% a structure at a ~ so this keeps an author's name from being broken across
% two lines.
% use \thanks{} to gain access to the first footnote area
% a separate \thanks must be used for each paragraph as LaTeX2e's \thanks
% was not built to handle multiple paragraphs
%

\author{Adrien~M.~Bizeray,~\IEEEmembership{}
        Jin--Ho~Kim,
        Stephen~R.~Duncan,~\IEEEmembership{Member,~IEEE,}
        and~David~A.~Howey,~\IEEEmembership{Senior~Member,~IEEE}% <-this % stops a space
\thanks{J--H Kim is with the Energy Lab, Samsung Advanced Institute of Technology, Samsung Electronics, 130 Samsung-ro, Suwon-si, 443-803, Korea (e-mail: jh527.kim@samsung.com).}
\thanks{A. M. Bizeray, S. R. Duncan and D. A. Howey are with the Energy and Power Group, Department of
Engineering Science, University of Oxford, Oxford, OX1 3PJ, U.K. (e-mail: adrien.bizeray@eng.ox.ac.uk; stephen.duncan@eng.ox.ac.uk; david.howey@eng.ox.ac.uk).}
\thanks{Manuscript received XXXX; revised XXXX.}}

% note the % following the last \IEEEmembership and also \thanks - 
% these prevent an unwanted space from occurring between the last author name
% and the end of the author line. i.e., if you had this:
% 
% \author{....lastname \thanks{...} \thanks{...} }
%                     ^------------^------------^----Do not want these spaces!
%
% a space would be appended to the last name and could cause every name on that
% line to be shifted left slightly. This is one of those "LaTeX things". For
% instance, "\textbf{A} \textbf{B}" will typeset as "A B" not "AB". To get
% "AB" then you have to do: "\textbf{A}\textbf{B}"
% \thanks is no different in this regard, so shield the last } of each \thanks
% that ends a line with a % and do not let a space in before the next \thanks.
% Spaces after \IEEEmembership other than the last one are OK (and needed) as
% you are supposed to have spaces between the names. For what it is worth,
% this is a minor point as most people would not even notice if the said evil
% space somehow managed to creep in.

% The paper headers
\markboth{}%
{Shell \MakeLowercase{\textit{et al.}}: Bare Demo of IEEEtran.cls for Journals}
% The only time the second header will appear is for the odd numbered pages
% after the title page when using the twoside option.
% 
% *** Note that you probably will NOT want to include the author's ***
% *** name in the headers of peer review papers.                   ***
% You can use \ifCLASSOPTIONpeerreview for conditional compilation here if
% you desire.

% If you want to put a publisher's ID mark on the page you can do it like
% this:
%\IEEEpubid{0000--0000/00\$00.00~\copyright~2007 IEEE}
% Remember, if you use this you must call \IEEEpubidadjcol in the second
% column for its text to clear the IEEEpubid mark.

% use for special paper notices
%\IEEEspecialpapernotice{(Invited Paper)}

% make the title area
\maketitle

\begin{abstract}
%\boldmath
% \blindtext[1]
This paper investigates the identifiability and estimation of the parameters of the single particle model (SPM) for lithium-ion battery simulation. Identifiability is addressed both in principle and in practice. The approach begins by grouping parameters and partially non-dimensionalising the SPM to determine the maximum expected degrees of freedom in the problem. We discover that, excluding open circuit voltage, there are only six independent parameters. We then examine the structural identifiability by considering whether the transfer function of the linearised SPM is unique. It is found that the model is unique provided that the electrode open circuit voltage functions have a known non-zero gradient, the parameters are ordered, and the electrode kinetics are lumped into a single charge transfer resistance parameter. We then demonstrate the practical estimation of model parameters from measured frequency-domain experimental electrochemical impedance spectroscopy (EIS) data, and show additionally that the parametrised model provides good predictive capabilities in the time domain, exhibiting a maximum voltage error of 20 mV between model and experiment over a 10 minute dynamic discharge.
\end{abstract}
% IEEEtran.cls defaults to using nonbold math in the Abstract.
% This preserves the distinction between vectors and scalars. However,
% if the journal you are submitting to favors bold math in the abstract,
% then you can use LaTeX's standard command \boldmath at the very start
% of the abstract to achieve this. Many IEEE journals frown on math
% in the abstract anyway.

% Note that keywords are not normally used for peerreview papers.
% \begin{IEEEkeywords}
% IEEEtran, journal, \LaTeX, paper, template.
% \end{IEEEkeywords}

% For peer review papers, you can put extra information on the cover
% page as needed:
% \ifCLASSOPTIONpeerreview
% \begin{center} \bfseries EDICS Category: 3-BBND \end{center}
% \fi
%
% For peerreview papers, this IEEEtran command inserts a page break and
% creates the second title. It will be ignored for other modes.
\IEEEpeerreviewmaketitle

\section*{Nomenclature}
\subsection*{Roman}
\begin{IEEEdescription}[\IEEEusemathlabelsep\IEEEsetlabelwidth{$cccccc$}]
    \item[$\mathcal{A}$] Electrode surface area, \SI{}{m^2}
    \item[$a_i$] Specific active surface area, \SI{}{m^{-1}}
    \item[$c_e$] Electrolyte lithium concentration, \SI{}{mol.m^{-3}}
    \item[$c_i$] Lithium concentration (active material), \SI{}{mol.m^{-3}}
    \item[$c_i^0$] Initial lithium concentration, \SI{}{mol.m^{-3}}
    \item[$c_i^{max}$] Maximum lithium concentration, \SI{}{mol.m^{-3}}
    \item[$c_i^s$] Surface lithium concentration, \SI{}{mol.m^{-3}}
    \item[$D_i$] Electrode active material diffusivity, \SI{}{m^2.s^{-1}}
    \item[$\mathcal{F}$] Faraday's constant, \SI{}{C.mol^{-1}}
    \item[$H^0$] SPM transfer function, \SI{}{\ohm}
    \item[$H_i^d$] Diffusion model transfer function, \SI{}{A^{-1}}
    \item[$I$] Applied current, \SI{}{\ampere}
    \item[$i_{0,i}$] Exchange current density, \SI{}{A.m^{-2}}
    \item[$j_i$] Reaction rate, \SI{}{mol.m^{-2}.s^{-1}}
    \item[$k_i$] Reaction rate constant, \SI{}{m^{2.5}.mol^{-0.5}.s^{-1}}
    \item[$L_j$] Single DoD loss function, \SI{}{\ohm^2}
    \item[$L$] Combined DoDs loss function, \SI{}{\ohm^2}
    \item[$N_\omega$] Number of data points per EIS measurement
    \item[$N_{DoD}$] Number of EIS datasets at different DoDs
    \item[$Q$] Charge/discharge capacity, \SI{}{\coulomb}
    \item[$Q_i^{th}$] Theoretical electrode capacity, \SI{}{\coulomb}.
    \item[$R$] Gas constant, \SI{}{J.mol^{-1}.K^{-1}}
    \item[$R_{ct}^0$] Charge-transfer resistance, \SI{}{\ohm}
    \item[$R_i$] Active material particle radius, \SI{}{\meter}
    \item[$r_i$] Particle radial coordinates, \SI{}{\meter}
    \item[$\bar{r}$] Particle dimensionless radial coordinates
    \item[$T$] Temperature, \SI{}{\kelvin}
    \item[$t$] Time, \SI{}{\second}
    \item[$U_i$] Open-circuit voltage, \SI{}{\volt}
    \item[$\bar{u}_i$] Change of variable $\bar{r}\bar{x}_i$
    \item[$V$] Cell terminal voltage, \SI{}{\volt}
    \item[$\bar{V}^0$] Voltage deviation from equilibrium, \SI{}{\volt}
    \item[$V_i$] Volume of electrode active material, \SI{}{m^3}
    \item[$\vect{x}_0$] Linearisation reference point
    \item[$x_i$] Lithium stoichiometry (active material)
    \item[$x_i^0$] Initial lithium stoichiometry
    \item[$x_i^s$] Surface lithium stoichiometry
    \item[$\bar{x}_i$] Change of variable $x_i - x_i^0$
    \item[$Z_j$] Measured cell impedance, \SI{}{\ohm}
\end{IEEEdescription}

\subsection*{Greek}
\begin{IEEEdescription}[\IEEEusemathlabelsep\IEEEsetlabelwidth{$cccccc$}]
    \item[$\alpha_i$] Anodic charge-transfer coefficient
    \item[$\alpha_i^0$] OCV gradient w.r.t stoichiometry, \SI{}{V}
    \item[$\beta_i^0$] OCV gradient w.r.t capacity, \SI{}{V/(A.s)}
    \item[$\delta_i$] Electrode thickness, \SI{}{\meter}
    \item[$\epsilon_i$] Solid-phase volume fraction
    \item[$\eta_i$] Overpotential, \SI{}{\volt}
    \item[$\theta_j$] Grouped parameter
    \item[$\greekvect{\theta}$] Vector of grouped parameters, $\greekvect{\theta} \in \mathbb{R}^6$
    \item[$\greekvect{\tilde{\theta}}$] Vector of identifiable parameters, $\greekvect{\tilde{\theta}} \in \mathbb{R}^3$
    \item[$\greekvect{\theta}_{id}$] Vector of identified parameters
    \item[$\tau_i^d$] Diffusion time constant, \SI{}{\second}
    \item[$\tau_i^k$] Kinetic time constant, \SI{}{\second}
    \item[$\omega$] Angular frequency, \SI{}{rad.s^{-1}}
\end{IEEEdescription}

% \subsection*{Abbreviations}
% \begin{IEEEdescription}[\IEEEusemathlabelsep\IEEEsetlabelwidth{$cccccc$}]
%     \item[BMS] Battery Management System
%     \item[DoD] Depth-of-Discharge
%     \item[ECM] Equivalent Circuit Model
%     \item[EIS] Electrochemical Impedance Spectroscopy
%     \item[LCO] Lithium Cobalt Oxide
%     \item[NMC] Nickel Manganese Cobalt oxide 
%     \item[OCV] Open Circuit Voltage
%     \item[P2D] Pseudo-Two Dimensional
%     \item[RMS] Root Mean Square
%     \item[SoC] State-of-Charge
%     \item[SoH] State-of-Health
%     \item[SPM] Single Particle Model
% \end{IEEEdescription}

\section{Introduction}
In addition to their widespread use in consumer electronics, the importance of lithium-ion batteries is increasing in large-scale energy storage applications, such as automotive, off-grid and grid applications. Such large-scale battery energy storage systems are required to operate for a decade or more. Battery degradation, which results in capacity and power fade over time, becomes a particularly important issue in applications with such long service life. However, battery management system (BMS) technologies, currently used in commercial applications for monitoring and controlling batteries to ensure their safety, have very limited insights into battery degradation.

Two important functions of a conventional BMS are (i) to monitor the battery voltage, current and temperature and ensure that these remain within their region of safe operation, and (ii) to estimate unmeasurable metrics such as State-of-Charge (SoC) and State-of-Health (SoH). The SoC denotes the remaining energy in the cell compared to a fully charged cell, while the SoH is usually defined as the capacity or resistance of an aged cell compared to that of a pristine cell.
A mathematical model of a battery is required to infer SoC and SoH from the available measurements of voltage, current and temperature. Conventional BMSs commonly employ electrical equivalent circuit models (ECMs) consisting of a voltage source and a network of resistors and capacitors whose current-voltage response mimics that of a battery. The parameters of such ECMs may vary with SoC, SoH, current and temperature and must therefore be identified from experimental data under a large range of operating conditions. However, the battery behaviour predicted by ECMs cannot credibly be extrapolated outside the range of validity over which they have been parametrised. Also, accounting for battery degradation is challenging due to the lack of physical significance of the model parameters \cite{Bizeray2015}.

\subsection*{Motivations}
In order to improve the lifetime performance of battery packs, researchers are investigating the application of physics-based electrochemical models instead of ECMs in the BMS, for example to enable electrochemical control of fast charging \cite{Klein2010,Moura2013a,Perez2014}. 
Due to their physical basis, such first-principle models can provide insights into battery ageing because they can more easily be coupled to degradation models. Physics-based models could be solved online in next-generation BMSs to enable health-aware management and control algorithms that use internal electrochemical safety limits (e.g.\ local overpotential or lithium concentration) instead of the commonly employed terminal voltage and current safety limits \cite{Chaturvedi2010, Moura2014}. 

However, this is a challenging research area from both the modelling and numerical solution perspective, and also the application perspective. One key challenge is whether such models are observable and can be used reliably for state estimation. Recent promising results indicate that electrochemical models could indeed be used in this way \cite{Santhanagopalan2006a,Smith2008b,DiDomenico2010,Smith2010,Moura2013b,Stetzel2015a,Bizeray2015,Zhao2015a}. Another important condition for the use of an electrochemical model in a BMS is that the parameters of the model must be credible, not just to recreate the measured voltage accurately, but so that internal states, such as overpotential, are also accurate. In general, parameters in these models are obtained from the literature (for example \cite{Northrop2011b,Bizeray2015}), and whilst this is useful for initial design studies of cell performance, for use in a BMS, the model must be parametrised specifically for the cells used in the pack. Relatively few authors have tackled this parameter estimation problem \cite{Schmidt2010,Ramadesigan2011,Forman2011c,Forman2012,Marcicki2013a} although it has been investigated in the context of ECMs of battery dynamics \cite{Jiang2011,Nazer2012,Jang2011,Moubayed2008}. The present work is therefore motivated by the need to obtain meaningful model parameters for using electrochemical battery models for control and estimation purposes.

\subsection*{Contributions}
In this paper we investigate the parameter estimation challenge in a simplified version of the `Doyle-Fuller-Newman' pseudo-two-dimensional (P2D) lithium-ion battery model \cite{Doyle1993}, the so-called single particle model (SPM) \cite{Atlung1979,Ning2004}, first to see whether groups of parameters are identifiable in principle, and then to investigate whether in practice they can be identified from simulated and experimental frequency-domain Electrochemical Impedance Spectroscopy (EIS) data \cite{Barsoukov2005}.

The main contribution of our work is to show the importance of considering the structural and practical identifiability of model parameters before undertaking any parameter estimation. Attempting the identification of all model parameters simultaneously without considering their identifiability is likely to lead to ill-conditioned optimisation problems. An electrochemical model identified in such a way may indeed fit the battery voltage and current response well, but the estimated parameters will have no physical significance and therefore the internal states cannot be reliably used for control. We also demonstrate parameter estimation of an electrochemical battery model from frequency-domain data obtained from commercial cells, and this is a relatively new and unexplored research topic \cite{Tippman2014,Murbach2017}. 

Our first key result is that the parameter identifiability of any lithium-ion battery model, whether ECM or first-principle electrochemical model, is largely conditional on the slope of each electrodes' open-circuit voltage (OCV) as a function of SoC. A flat OCV curve overshadows all dynamics and results in parameter unidentifiability. A second key message from this work is that physics-based models for lithium-ion batteries are over-parametrised and therefore one must try to identify the minimum number of identifiable grouped parameters prior to attempting any parameter estimation.

\section{The single-particle model}
The single particle model (SPM) was first introduced by \cite{Atlung1979} and later extended to lithium-ion batteries by \cite{Ning2004}. The SPM is an electrochemical battery model describing the cell thermodynamics, the diffusion of lithium in the active material of the electrodes, and the interfacial kinetics of the lithium intercalation/de-intercalation reaction at the electrode/electrolyte interface. In contrast with the more complex P2D model \cite{Doyle1993}, electrolyte dynamics are neglected in the SPM and therefore the reaction rate is assumed uniform for each electrode. We chose the SPM in this paper because it is the `simplest' electrochemical lithium-ion model and a good starting point for parameter identification studies. We anticipate that our methodology could be extended to other electrochemical models suitable for higher C-rates. Importantly, we do not anticipate that our main conclusions will change with respect to more complex models: the electrode OCV functions will remain key in paramaterisation, and the over-parametrisation issue will become more challenging. Although the validity range of the SPM may vary depending on the physico-chemical properties of the cell (e.g.\ electrolyte conductivity, porosity), it is commonly accepted in the literature that the SPM is valid at currents below 1\textit{C} to 2\textit{C}\footnote{In the battery literature the electrical current is commonly expressed as a \textit{C}-rate, which is a value of current normalised by the cell capacity so that a current amplitude of 1\text{C} will discharge the cell in one hour.}, where concentration gradients in the electrolyte are negligible \cite{Chaturvedi2010, Schmidt2010,Santhanagopalan2007}. For many applications, including electric vehicles and grid storage, average C-rates are lower than 1\textit{C} and this limitation is not particularly restrictive. In fact, we demonstrate later in this paper that reasonable voltage errors may be obtained with the \textit{linearised} SPM on data including peaks in current of up to 6\textit{C}. Finally, for the purpose of parameter estimation, in our work we used frequency-domain linear response EIS experimental data. Linear EIS data are recorded by applying a small AC current signal (with zero DC offset) to the cell with an amplitude much lower than 1\textit{C} (typically around \textit{C}/10) to ensure that the cell voltage response remains linear. It is therefore reasonable to use the SPM for modelling the battery impedance from linear EIS data with such low amplitude currents. 

\subsection{Spherical particle diffusion model}
In the SPM, the diffusion of lithium in the active material of each battery electrode $i$ (anode or cathode) is governed by the Fickian diffusion equation in spherical coordinates
	\begin{equation}
		\frac{\partial c_i}{\partial t} = \frac{D_i}{r_i^2} \frac{\partial}{\partial r_i} \left(  r_i^2  \frac{\partial c_i}{\partial r_i} \right)  ,
		\label{eq:diffusionEquation}
	\end{equation}
where $r_i$ denotes the radial coordinates, $c_i$ the lithium concentration profile and $D_i$ the lithium diffusion coefficient (assumed uniform and constant) in electrode $i$. The subscript $i$ either takes the value $+$ or $-$ to refer to the cathode or anode domain respectively. The diffusion equation is subject to Neumann boundary conditions at the particle centre $r_i=0$ and surface $r_i=R_i$,
	\begin{equation}
		\left. \frac{\partial c_i}{\partial r_i} \right|_{r_i=0} = 0
		\quad \text{and} \quad
		D_i \left. \frac{\partial c_i}{\partial r_i} \right|_{r=R_i} = -j_i  ,
		\label{eq:boundaryConditions}
	\end{equation}
and the initial condition equation,
	\begin{equation}
		c_i (0,r_i) = c_i^0,
		\label{eq:initialConditions}
	\end{equation}
where $c_i^0$ is the initial concentration profile in each particle.
The molar flux $j_i$ in \eqref{eq:boundaryConditions} expresses the rate of the lithium intercalation/de-intercalation reaction.
In contrast to the Doyle--Fuller--Newman model \cite{Doyle1993} where algebraic equations must be solved to obtain the molar flux for each particle, the molar fluxes in the SPM can be directly related to the applied battery current $I$ through the relationships,
	\begin{equation}
		\label{eq:current_reactionRate_relation}
		j_{-} = \frac{+I}{a_- \delta_{-} \mathcal{F}\mathcal{A}} 	
		\quad \text{and} \quad 
		j_{+} = \frac{-I}{a_+ \delta_{+}\mathcal{F}\mathcal{A}}  ,
	\end{equation}
where $a_i = 3\epsilon_i/R_i$ is the specific active surface area in electrode $i$, $\epsilon_i$ is the volume fraction of active material in an electrode, $\mathcal{F}$ is Faraday's constant, $\mathcal{A}$ is the electrode surface area which is assumed equal for both electrodes, and $\delta_i$ is the thickness of the electrode $i$. 
By convention, the battery discharge current is positive and the charging current is negative. The anode is the negative electrode upon discharge and the cathode is the positive electrode. Although strictly speaking these swap upon charging, we will in this paper use the widely accepted abuse of terminology that the `anode' is the negative electrode and the `cathode' is the positive electrode.
The above molar flux expressions assume an electrochemical reaction involving a single electron transfer, which is the case for the lithium intercalation/de-intercalation reaction.

\subsection{Voltage measurement equation}
The initial-boundary value problems~(\ref{eq:diffusionEquation}-\ref{eq:initialConditions}) for each electrode constitute the dynamic part of the single-particle model.
The battery terminal voltage~$V$ is given by the nonlinear measurement equation,
	\begin{equation}
		\label{eq:voltage}
		V = U_+ \left( x_+^s \right) - U_- \left( x_-^s \right) + \eta_+ - \eta_- .
	\end{equation}
The anode and cathode OCV $U_-$ and $U_+$ respectively are empirical nonlinear functions of the surface stoichiometry $x_i^s$ of each particle defined according to,
    \begin{equation}
    x_{i}^s = c_i^s/c_i^{max}.
    \end{equation} 
The overpotential  $\eta_i$ is the voltage drop due to the departure from equilibrium potential associated with the intercalation/de-Â­intercalation reaction in each electrode. The relationship between the reaction rate $j_i$ and the overpotential $\eta_i$ is given by the ButlerÂ­-Volmer kinetics equation \cite{Doyle1993}
	\begin{equation}
		\label{eq:butler-volmer}
		j_i = \frac{i_{0,i}}{\mathcal{F}} \left[ \exp \left( \frac{\alpha_i \mathcal{F}}{RT} \eta_i \right) - \exp \left( \frac{ -\left(1-\alpha_i \right) \mathcal{F}}{RT} \eta_i \right) \right].
	\end{equation}
The exchange current density $i_{0,i}$ depends on the reaction rate constant $k_i$ and the reactants and products concentrations, i.e.\ the electrolyte concentration $c_e$ (assumed constant in this model), and the lithium solid-phase concentration $c_i^s$ at the surface of the particle, through the relation
	\begin{equation}
		i_{0,i} = k_i \mathcal{F} \sqrt{c_e} \sqrt{c_i^s} \sqrt{c_i^{max} - c_i^s }.
	\end{equation}
By assuming that the anodic charge transfer coefficients in~\eqref{eq:butler-volmer} are $\alpha_i = 0.5$, i.e.\ anodic and cathodic charge transfer coefficients are equal, the overpotential $\eta_i$ can be expressed as a function of the reaction rate $j_i$ by
	\begin{equation}
		\label{eq:overpotential}
		\eta_i = \frac{2RT}{\mathcal{F}} \sinh^{-1} \left( \frac{j_i \mathcal{F} }{2 i_{0,i}} \right).
	\end{equation}

\subsection{Identification of grouped parameters}
First-principle models, such as the SPM, are usually over-parametrised in the sense that only a subset of parameters can be estimated from measured input-output data. A first necessary step towards credible parameter estimation is to reformulate the model in terms of the minimum number of parameter groups.
In order to identify such groups of parameters, we first introduce the dimensionless radial coordinates $\bar{r} = r_i / R_i$, and the stoichiometry $x_i = c_i/c_i^{max}$,  instead of the concentration in the governing equations.
In addition, the change of variable $\bar{x}_i = x_i - x_i^0$ is introduced, with $x_i^0 = c_i^0/c_i^{max}$ being the initial stoichiometry in each particle, in order to set the initial condition to zero in the governing equations.
Note that the time independent variable $t$ is kept dimensional for simplicity because the diffusion time constants are different in each electrode. 
Introducing these dimensionless variables, the initial-boundary value problem~(\ref{eq:diffusionEquation}-\ref{eq:initialConditions}) can be written as
	\begin{equation}
		\frac{\partial \bar{x}_i}{\partial t} =  
		\frac{D_i}{R_i^2} \frac{1}{\bar{r}^2} \frac{\partial }{\partial \bar{r}} \left( \bar{r}^2 \frac{\partial \bar{x}_i}{\partial \bar{r}}\right) ,
	\end{equation}
subject to the boundary conditions
	\begin{equation}
		\left. \frac{\partial \bar{x}_i}{\partial \bar{r}} \right|_{\bar{r}=0} = 0  
		\quad \text{and} \quad 
		\left. \frac{\partial \bar{x}_i}{\partial \bar{r}} \right|_{\bar{r}=1} = \frac{- R_i}{D_i c_i^{max}} j_i ,
	\end{equation}
with initial conditions
	\begin{equation}
		\bar{x}_i(0,\bar{r})=0.
	\end{equation}
The boundary condition at the surface of the particle in each electrode can also be written in terms of the input current $I$ instead of the fluxes $j_i$ using relations~\eqref{eq:current_reactionRate_relation} according to
	\begin{align}
		\left. \frac{\partial \bar{x}_+}{\partial \bar{r}} \right|_{\bar{r}=1} &= + \frac{R_+^2}{D_+} \frac{I}{3 \epsilon_+  \delta_{+} c_+^{max}\mathcal{F}\mathcal{A}}
		\\
		\left. \frac{\partial \bar{x}_-}{\partial \bar{r}} \right|_{\bar{r}=1} &= - \frac{R_-^2}{D_-} \frac{I}{3 \epsilon_-\delta_{-} c_-^{max}\mathcal{F}\mathcal{A}}.
	\end{align}
Similarly using relation~\eqref{eq:current_reactionRate_relation} the expression for the overpotential in each electrode~\eqref{eq:overpotential} becomes 
	\begin{align}
		\eta_+ &= \frac{2RT}{\mathcal{F}} \sinh^{-1} \left( - \frac{R_+}{2k_+ \sqrt{c_e}} \frac{1}{3 \epsilon_+  \delta_{+} c_+^{max}\mathcal{F}\mathcal{A}} \frac{I}{\sqrt{x_+^s (1-x_+^s)}} \right)
		\\
		\eta_- &= \frac{2RT}{\mathcal{F}} \sinh^{-1} \left( + \frac{R_-}{2k_- \sqrt{c_e}} \frac{1}{3 \epsilon_-\delta_{-} c_-^{max}\mathcal{F}\mathcal{A}} \frac{I}{\sqrt{x_-^s (1-x_-^s)}} \right)
	\end{align}
with the voltage measurement~\eqref{eq:voltage} expression unchanged.
Six physically meaningful groups of parameters naturally arise in these equations. Indeed one can identify three groups of parameters for each electrode: a diffusion time constant $\tau^d_i$, kinetics time constant $\tau^k_i$ and the maximum theoretical electrode capacity $Q^{th}_i$. The expressions for these six physically meaningful groups of parameters are defined as follows
	\begin{align}
		\label{eq:diffusionTimeCste}
		\tau_+^d &= \frac{R_+^2}{D_+} & 
		\tau_-^d &= \frac{R_-^2}{D_-} 
		\\
		\label{eq:kineticsTimeCste}
		\tau_+^k &= \frac{R_+}{2 k_+ \sqrt{c_e}} & 
		\tau_-^k &= \frac{R_-}{2 k_- \sqrt{c_e}} 
		\\
		\label{eq:thCapacities}
		Q_+^{th} &= - \epsilon_+ \delta_+ c_+^{max} \mathcal{F}\mathcal{A} & 
		Q_-^{th} &= + \epsilon_- \delta_- c_-^{max} \mathcal{F}\mathcal{A}
	\end{align}
Note that for convenience the cathode theoretical capacity $Q_+^{th}$ is defined negative by convention and without loss of generality, to yield the same model structure in both the anode and cathode.
Substituting these grouped parameters into the governing equations results in the following diffusion equation for both the cathode and anode:
	\begin{equation}
		\label{eq:diffEq_groups}
		\frac{\partial \bar{x}_i}{\partial t} =  \frac{1}{\tau_i^d} \frac{1}{\bar{r}^2} \frac{\partial }{\partial \bar{r}} \left( \bar{r}^2 \frac{\partial \bar{x}_i}{\partial \bar{r}}\right)
	\end{equation}
subject to the boundary conditions
	\begin{equation}
		\label{eq:bcEq_groups}
		\left. \frac{\partial \bar{x}_i}{\partial \bar{r}} \right|_{\bar{r}=0} = 0  
		\quad \text{and} \quad 
		\left. \frac{\partial \bar{x}_i}{\partial \bar{r}} \right|_{\bar{r}=1} = - \frac{\tau^d_i}{3Q^{th}_i} I,
	\end{equation}
with initial condition
	\begin{equation}
		\label{eq:initEq_groups}
		\bar{x}_i(0,\bar{r}) = 0 .
	\end{equation}
The voltage measurement equation remains 
	\begin{equation}
		\label{eq:voltageMeasurementGroups}
		V = U_+ \left( x_+^s \right) - U_- \left( x_-^s \right) + \eta_+ - \eta_- ,
	\end{equation}
with the cathode and anode overpotentials given by 
	\begin{align}
		\label{eq:cathodeOverpotentialGroups}
		\eta_+ &= \frac{2RT}{F} \sinh^{-1} \left(  \frac{\tau_+^k}{3Q_+^{th}} \frac{I}{\sqrt{x_+^s (1-x_+^s)}}\right)
		\\
		\label{eq:anodeOverpotentialGroups}
		\eta_- &= \frac{2RT}{F}  \sinh^{-1} \left( \frac{\tau_-^k}{3Q_-^{th}} \frac{I}{\sqrt{x_-^s (1-x_-^s)}} \right).
	\end{align}
Inspecting \eqref{eq:diffEq_groups} through \eqref{eq:anodeOverpotentialGroups}, one can identify six groups of parameters $\theta_j$ that fully parametrise the SPM given by the parameter vector $\greekvect{\theta} \in \mathbb{R}^6$:
	\begin{equation}
	    \label{eq:fullParameterDef}
		\greekvect{\theta} = \left[ 
		\tau_+^d \quad
		\frac{\tau_+^d}{3 Q_+^{th}} \quad
		\frac{\tau_+^k}{3 Q_+^{th}} \quad
		\tau_-^d \quad
		\frac{\tau_-^d}{3 Q_-^{th}} \quad
		\frac{\tau_-^k}{3 Q_-^{th}} \quad
		\right]^T .
	\end{equation}
Unlike the parameters defined in~(\ref{eq:diffusionTimeCste}-\ref{eq:thCapacities}), these parameters $\theta_j$ appear only once and not as a product of each other in the model equations. Moreover, one can show that there is a one-to-one mapping between the six parameters $\theta_j$ and the six parameters \eqref{eq:diffusionTimeCste}--\eqref{eq:kineticsTimeCste}--\eqref{eq:thCapacities} defined by:
	\begin{equation}
	    \label{eq:parameterMapping}
		\begin{cases}
			\tau_+^d 	&= \theta_1 \\
			\tau_+^k  	&= (\theta_1\theta_3)/ \theta_2 \\
			Q_+^{th} 	&= \theta_1/(3 \theta_2) \\
			\tau_-^d 	&= \theta_4 \\
			\tau_-^k 	&= ( \theta_4 \theta_6)/ \theta_5 \\
			Q_-^{th} 	&= \theta_4 / (3 \theta_5)
		\end{cases}
	\end{equation}
Therefore, assuming that the initial electrode stoichiometries $x_i^0$ (i.e.\ state-of-charge) are known and that the OCVs are known functions of the surface stoichiometry for each electrode, the six parameters contained in $\greekvect{\theta}$, or equivalently the six parameters defined in (\ref{eq:diffusionTimeCste}--\ref{eq:thCapacities}), are sufficient to fully parametrise the SPM.

\section{Structural identifiability}
\label{sec:structural_identifiability}
Six grouped parameters have been identified as sufficient to fully parametrise the SPM. However, this does not imply that these six parameters can be identified from the battery current-voltage response. Several approaches can be used to investigate the parameter identifiability of a dynamical model from input-output data. In this section, the so-called \emph{structural identifiability} \cite{Bellman1970} of the SPM is discussed. Structural identifiability investigates the mathematical identifiability of the model irrespective of the identification data considered (which are also assumed noise-free). A definition of structural identifiability for linear time-invariant dynamic models that can be cast into a transfer function $H(s,\greekvect{\theta})$ parametrised by a vector of parameters $\greekvect{\theta}$ is given as follows~\cite{Ljung1999,Alavi2016}:
\begin{Definition}\label{def:identifiability}
Consider a model structure $\mathcal{M}$ with the transfer function $H(s,\greekvect{\theta})$ parametrised by $\greekvect{\theta} \in \mathcal{D}\subset \mathbb{R}^n$ where $n$ denotes the number of parameters of the model.
The identifiability equation for $\mathcal{M}$ is given by 
	\begin{equation}
		\label{TFd:ide}
		H(s,\greekvect{\theta})=H(s,\greekvect{\theta}^\ast)\qquad \text{for almost all } s,
	\end{equation}
where $\greekvect{\theta}, \greekvect{\theta}^\ast\in\mathcal{D}$. The model structure $\mathcal{M}$ is said to be
	\begin{itemize}
		\item \emph{globally identifiable} if \eqref{TFd:ide} has a unique solution in $\mathcal{D}$,
		\item \emph{locally identifiable} if \eqref{TFd:ide} has a finite number of solutions in $\mathcal{D}$,
		\item \emph{unidentifiable} if \eqref{TFd:ide} has a infinite number of solutions in $\mathcal{D}$.
	\end{itemize}
\end{Definition}

The SPM is not a linear dynamical model because of its voltage measurement equation \eqref{eq:voltageMeasurementGroups} incorporating nonlinear Butler-Volmer kinetics (\ref{eq:cathodeOverpotentialGroups},\ref{eq:anodeOverpotentialGroups}), and nonlinear OCV functions. Therefore Definition~\ref{def:identifiability} cannot directly be used to investigate the SPM structural identifiability unless we first linearise the model by assuming a small perturbation around a fixed depth-of-discharge (DoD) point. Since we assume the nonlinear OCV functions are measured and known a priori (for example using the techniques described in \cite{Birkl2016,McTurk2015}), linearisation is a valid approach to obtain the parameters of the diffusion sub-models, which are linear, and to obtain a linearised approximation of the kinetics. The subject of identifiability and parameter estimation of the nonlinear kinetics term in the output equation would be an interesting topic for further research.

\subsection{Diffusion model transcendental transfer function}
Definition~\ref{def:identifiability} requires the model transfer function; we therefore first derive this for the spherical particle diffusion model and then combine it in subsequent sections with the transfer function of the linearised voltage measurement equation to derive the transfer function for the linearised SPM.
The transfer function for the spherical diffusion model has previously been derived in the literature for determining the diffusion coefficient of various electrode materials in a laboratory experimental setup \cite{Haran1998,Guo2002} and studying the cell frequency response analytically \cite{Sikha2007}, but is included here for completeness.
In order to simplify the derivation, the change of variable $\bar{u}_i = \bar{r} \bar{x}_i$ is introduced into the initial-boundary value problem (\ref{eq:diffEq_groups}-\ref{eq:initEq_groups}) describing the diffusion in a spherical particle. Under this change of variable \eqref{eq:diffEq_groups} can be equivalently written
	\begin{equation}
		\label{eq:diffEq_groups_variableChange}
		\frac{\partial \bar{u}_i}{\partial t} = \frac{1}{\tau_i^d} \frac{\partial^2 \bar{u}_i}{\partial \bar{r}^2},
	\end{equation}
and the boundary conditions \eqref{eq:bcEq_groups} become
	\begin{equation}
		\label{eq:bcEq_groups_variableChange}
		\bar{u}_i \left(\bar{r} = 0 \right) = 0 
		\quad \text{and} \quad 
		\left.\frac{\partial \bar{u}_i}{\partial \bar{r}}\right|_{\bar{r}=1} - \bar{u}_i(\bar{r}=1) = \frac{- \tau_i^d}{3 Q_i^{th}} I.
	\end{equation}
Note that the homogeneous Neumann boundary condition at the centre of the particle can be reduced to a simpler homogeneous Dirichlet boundary condition so that $\lim_{r_i \rightarrow 0} c_i(r_i)$ remains finite.
By introducing this change of variable the initial condition~\eqref{eq:initEq_groups} conveniently becomes
	\begin{equation}
		\label{eq:initEq_groups_variableChange}
		\bar{u}_i (0,\bar{r}) = 0.
	\end{equation}

Since the initial-boundary value diffusion problem \eqref{eq:diffEq_groups_variableChange}--\eqref{eq:bcEq_groups_variableChange}--\eqref{eq:initEq_groups_variableChange} is linear, an equivalent transfer function can be determined without loss of generality.
 Taking the Laplace transform of~\eqref{eq:diffEq_groups_variableChange} yields
	\begin{equation}
		\frac{d^2 \bar{U}_i(s,\bar{r})}{d\bar{r}^2} - s \tau_i^d \bar{U}_i(s,\bar{r}) = 0
	\end{equation}
where $s$ is the frequency-domain Laplace variable.
The characteristic equation for this differential equation is 
	\begin{equation}
		\lambda^2 - s \tau_i^d = 0 \quad \Rightarrow \quad \lambda = \pm \sqrt{s \tau_i^d}, 
	\end{equation}
and its general solution is therefore
	\begin{equation}
		\label{eq:generalSolution1}
		\bar{U}_i(s,\bar{r}) = A_i(s) e^{+\bar{r} \sqrt{s \tau_i^d}} + B_i(s) e^{-\bar{r} \sqrt{s \tau_i^d} }
	\end{equation}
with $A_i(s)$, $B_i(s)$ two constants (with respect to $\bar{r}$) to be determined using the boundary conditions~\eqref{eq:bcEq_groups_variableChange}.
Substituting~\eqref{eq:generalSolution1} into the particle centre boundary condition~\eqref{eq:bcEq_groups_variableChange} at $\bar{r} = 0$ yields
	\begin{equation}
		\label{eq:constantB}
		B_i(s) = -A_i(s).
	\end{equation}
And substituting~\eqref{eq:generalSolution1} and~\eqref{eq:constantB} into the surface boundary conditions ~\eqref{eq:bcEq_groups_variableChange} at $\bar{r}=1$ yields
	\begin{equation}
		\label{eq:constantA}
		A_i(s) = \frac{\tau_i^d}{3 Q_i^{th}} \frac{I(s)/2}{\sinh{(\sqrt{s \tau_i^d)}} - \sqrt{s \tau_i^d} \cosh{(\sqrt{s \tau_i^d)}}}
	\end{equation}
Substituting~\eqref{eq:constantA}--\eqref{eq:constantB} into~\eqref{eq:generalSolution1} yields the general solution
	\begin{equation}
		\label{eq:generalSolution2}
		\bar{U}_i(s,\bar{r}) = \frac{\tau_i^d}{3 Q_i^{th}} \frac{ \sinh{(\bar{r} \sqrt{s \tau_i^d})} }{ \sinh{(\sqrt{s \tau_i^d)}} - \sqrt{s \tau_i^d} \cosh{(\sqrt{s \tau_i^d)}} } I(s).
	\end{equation}
The variable of interest is the surface stoichiometry $\bar{X}^s_i(s)$, rather than $\bar{U}_i(s,\bar{r})$, since this is the variable involved in the voltage measurement equation. By substituting $\bar{u}_i = \bar{r} \bar{x}_i$ into~\eqref{eq:generalSolution2}, evaluating at $\bar{r}=1$ and dividing by the input current $I(s)$, the transfer function $H_i^d(s)$ from current to surface stoichiometry for the spherical diffusion model is given by
	\begin{equation}
		H_i^d(s) = \frac{\bar{X}_i^s(s)}{I(s)} =  \frac{\tau_i^d}{3 Q_i^{th}}  \frac{ \tanh{(\sqrt{s \tau_i^d})} }{ \tanh{(\sqrt{s \tau_i^d)}} - \sqrt{s \tau_i^d} } .
	\end{equation}
The cathode and anode diffusion transfer functions expressed in terms of the parameter vector $\greekvect{\theta}$ are therefore respectively
	\begin{equation}
		\label{eq:cathodeTF}
		H_+^d (s,\greekvect{\theta}) =  \frac{ \theta_2 \tanh{(\sqrt{s \theta_1}}) }{ \tanh{(\sqrt{s \theta_1)}} - \sqrt{s \theta_1} },
	\end{equation}
and
	\begin{equation}
		\label{eq:anodeTF}
		H_-^d (s,\greekvect{\theta}) =  \frac{ \theta_5 \tanh{(\sqrt{s \theta_4}}) }{ \tanh{(\sqrt{s \theta_4)}} - \sqrt{s \theta_4} } .
	\end{equation}

\subsection{Linearisation of the voltage measurement equation}
The voltage measurement equation is a nonlinear function of the anode and cathode surface stoichiometry $x_-^s$ and $x_+^s$ respectively, and the input current $I$.
Assuming that the input current amplitude $I$ remains small and the battery is operated close to its initial DoD, i.e.\ the stoichiometry in both electrodes remains close to the initial value $x_i^0$, the voltage equation can be linearised using a first-order Taylor series approximation about the reference point $\vect{x_0} = \left(x_+^0,x_-^0,I_0 = 0 \right)$ according to
	\begin{equation}
	    \label{eq:linearisedMeasEq}
		\bar{V}^0
		\approx 
		\left.\frac{\partial V}{\partial x_+^s}\right|_{\vect{x_0}}\bar{x}_+^s +
		\left. \frac{\partial V}{\partial x_-^s}\right|_{\vect{x_0}} \bar{x}_-^s +
		\left. \frac{\partial V}{\partial I}\right|_{\vect{x_0}} I,
	\end{equation}
where $\bar{V}^0 = V -V(\vect{x_0})$ denotes the deviation of the voltage from the equilibrium voltage at the linearisation point,
\begin{equation}
    V(\vect{x_0}) = U_+(x_+^0) - U_-(x_-^0).
\end{equation}

The partial derivative of the voltage $V$ with respect to the input current $I$ evaluated at the reference point is given by
	\begin{align}
		\left. \frac{\partial V}{\partial I}\right|_{\vect{x_0}} {}={}&
		\frac{2RT}{\mathcal{F}} 
		\left(
			\frac{\theta_3}{\sqrt{(1-x_+^0)x_+^0} \sqrt{1+ \left( \theta_3 I_0\right)^2}} 
		\right. \nonumber \\ 
		&{-}\left. \frac{\theta_6}{\sqrt{(1-x_-^0) x_-^0} \sqrt{1+ \left( \theta_6 I_0  \right)^2}}
		\right).
	\end{align}
which simplifies to 
	\begin{align}
		\label{eq:partialDerivativeCurrent}
		\left. \frac{\partial V}{\partial I}\right|_{\vect{x_0}} &= 
		\frac{2RT}{\mathcal{F}} 
		\left(
			\frac{\theta_3}{\sqrt{(1-x_+^0)x_+^0}}
		-
			\frac{\theta_6}{\sqrt{(1-x_-^0) x_-^0}}
		\right)
	\end{align}
when substituting for $I_0 = 0$.
% This term can be interpreted as a charge transfer resistance and will therefore be denoted $R_{ct}^0$, resulting in the charge transfer voltage drop $\eta_{ct} =  - R_{ct}^0  I$ in the voltage equation.
%
The partial derivative with respect to the surface stoichiometry in the cathode and anode are given respectively by
	\begin{align}
		\left.\frac{\partial V}{\partial x_+^s}\right|_{\vect{x_0}} {}={}&
		+ \left. \frac{d U_+}{d x_+^s} \right|_{x_+^s=x_+^0} \nonumber \\
		&{-}
		I_0 \frac{RT}{\mathcal{F}} \frac{\theta_3 \left(1-2 x_+^0\right) }{\sqrt{1+ \frac{\left( \theta_3 I_0 \right)^2}{(1-x_+^0)x_+^0}}\left((1-x_+^0)x_+^0 \right)^{3/2}}
	\end{align}
and
	\begin{align}
		\left.\frac{\partial V}{\partial x_-^s}\right|_{\vect{x_0}} {}={}&
		- \left. \frac{d U_-}{d x_-^s} \right|_{x_-^s=x_-^0} \nonumber \\
		&{+}
		I_0 \frac{RT}{\mathcal{F}} \frac{\theta_6 \left(1-2 x_-^0\right)}{\sqrt{1+\frac{\left(\theta_6 I_0\right)^2}{(1-x_-^0)x_-^0}}\left((1-x_-^0)x_-^0 \right)^{3/2}}.
	\end{align}
Substituting for $I_0 = 0$ yields the simple expressions
	\begin{align}
	    \label{eq:cathodeOCPgradient}
		\left.\frac{\partial V}{\partial x_+^s}\right|_{\vect{x_0}} &= 
		+ \left. \frac{d U_+}{d x^s_+} \right|_{x^s_+=x_+^0} =
		 \alpha_+^0 \\
		 \label{eq:anodeOCPgradient}
		\left.\frac{\partial V}{\partial x_-^s}\right|_{\vect{x_0}}  &= 
		-\left. \frac{d U_-}{d x^s_-} \right|_{x^s_-=x_-^0} = 
		- \alpha_-^0
	\end{align}
where $\alpha_i^0$ denotes the gradients of the OCV functions in each electrode, with respect to stoichiometry, at the linearisation point.
Substituting the partial derivatives in (\ref{eq:linearisedMeasEq}) by their respective expression given by (\ref{eq:partialDerivativeCurrent}), (\ref{eq:cathodeOCPgradient}) and (\ref{eq:anodeOCPgradient}) therefore yields the linearised voltage measurement equation
	\begin{equation}
		\label{eq:linearisedVoltageMeas}
		\bar{V}^0 (t)=   
		\alpha_+^0 \bar{x}_+^s(t) -
		\alpha_-^0 \bar{x}_-^s(t) -
		R_{ct}^0(\greekvect{\theta}) I(t).
	\end{equation}
where $R_{ct}^0(\greekvect{\theta})$ denotes the charge transfer resistance arising from the intercalation/de-intercalation reaction kinetics and defined according to
\begin{equation}
    \label{eq:chargeTransferResistance}
    R_{ct}^0(\greekvect{\theta})
    = 
    - \frac{2RT}{\mathcal{F}} 
		\left(
			\frac{\theta_3}{\sqrt{(1-x_+^0)x_+^0}}
		-
			\frac{\theta_6}{\sqrt{(1-x_-^0) x_-^0}}
		\right).
\end{equation}

\subsection{Transfer function of the linearised SPM}
\label{subsec:linearisedTF}
Taking the Laplace transform of the linearised voltage equation~\eqref{eq:linearisedVoltageMeas} and dividing by the input current $I$ yields the transfer function $H^0(s,\greekvect{\theta})=\bar{V}^0(s)/I(s)$ of the linearised SPM about the equilibrium point $\vect{x_0} = \left(x_+^0,x_-^0,I_0 = 0 \right)$:
	\begin{equation}
		\label{eq:voltageTF}
		H^0(s,\greekvect{\theta}) =
		\alpha_+^0 H_+^d(s,\greekvect{\theta}) - 
		\alpha_-^0 H_-^d(s,\greekvect{\theta})
		- R_{ct}^0(\greekvect{\theta}).
	\end{equation}
Substituting for the anode and cathode diffusion transfer function $H_+^d$ and $H_-^d$ with their respective expressions~\eqref{eq:cathodeTF} and~\eqref{eq:anodeTF} gives
	\begin{align}
	    \label{eq:voltageTF_fullExpression}
		H^0(s,\greekvect{\theta}) {}={}&
		\frac{ \alpha_+^0 \theta_2 \tanh{(\sqrt{s \theta_1})} }{ \tanh{(\sqrt{s \theta_1)}} - \sqrt{s \theta_1} } \nonumber \\
		&{-} 
		\frac{ \alpha_-^0 \theta_5 \tanh{(\sqrt{s \theta_4})} }{ \tanh{(\sqrt{s \theta_4)}} - \sqrt{s \theta_4} } 
		- R_{ct}^0(\greekvect{\theta})
	\end{align}
One can see that only the difference between the parameters $\theta_3$ and $\theta_6$ describing the cathode and anode kinetics appears in the charge transfer resistance term $R_{ct}(\greekvect{\theta})$ given by \eqref{eq:chargeTransferResistance}. As a result, there are an infinite number of pairs $(\theta_3,\theta_6)$ that will yield the same transfer function and only the lumped parameter $R_{ct}$ can be estimated using the linearised model at a given DoD. We have therefore reduced the parameter space to five parameters by combining the cathode and anode kinetics into the charge-transfer resistance term $R_{ct}$.

This assumes that the OCV slopes $\alpha_+^0$ and $\alpha_-^0$ are known parameters, but these are not directly measurable, since in practice, OCV can only be measured with respect to capacity (units Ah or Coulombs), not against stoichiometry (which is non-dimensional). By recognising that the theoretical capacity of an electrode is given by $Q_i^{th}=c_i^{max} \mathcal{F} V_i$ with $V_i=\epsilon A \delta_i$ (the volume of active material in the electrode), one can easily show that a variation of stoichiometry $\delta x_i$ is proportional to a variation of charge/discharge capacity $\delta Q$ according to
	\begin{equation}
		\delta x_i = \frac{\delta c_i}{c^{max}_i} = \frac{\delta c_i \mathcal{F}V_i}{Q_i^{th}} = \frac{\delta Q}{Q_i^{th}}.
	\end{equation}
Therefore the derivative of the OCV with respect to $x_i$ can be related to its derivative with respect to charge/discharge capacity $Q$, denoted $\beta_i^0$, according to
	\begin{equation}
	    \label{eq:beta_i}
		\beta_i^0 = \frac{\alpha_i^0}{Q^{th}_i}= \frac{1}{Q^{th}_i}\left. \frac{d U_i}{d x_i^s} \right|_{x_i^{s,0}} =  \left.\frac{d U_i}{d Q}\right|_{x_i^{s,0}} .
	\end{equation}
Substituting $\alpha_i^0$ for $\beta_i^0$ using \eqref{eq:beta_i} in \eqref{eq:voltageTF_fullExpression} and recalling the expressions of $Q_i^{th}$ in terms of $\theta_j$ in \eqref{eq:parameterMapping} yields the transfer function
	\begin{equation}
		H^0(s,\greekvect{\theta}) = 
		\beta_+^0 f\left( s,\theta_1 \right)
		-
		\beta_-^0 f\left( s,\theta_4 \right)
		-
		R_{ct}^0,
    \end{equation}
where the function $f$ defined as
	\begin{equation}
		f \left( s,\theta_j \right) = \frac{1}{3}\frac{ \theta_j \tanh{(\sqrt{s \theta_j})} }{  \tanh{(\sqrt{s \theta_j})} - \sqrt{s \theta_j}}
	\end{equation}
is only parametrised by the electrode diffusion time constant $\theta_1$ or $\theta_4$ in the cathode or anode respectively.
The coefficients $\beta +^0$ and $\beta_-^0$ are the measurable and assumed known OCV gradients with respect to capacity in the cathode and anode respectively. This requires access to half-cell or reference electrode cell data, which may be obtained from commercially available cells as we have recently demonstrated \cite{Birkl2016,McTurk2015}. In the absence of individual electrode OCV data, it is not possible to parametrise the SPM directly, although it may be reasonable to use a 2-electrode OCV measurement combined with available OCV data from the literature for the graphite negative electrode to infer the positive electrode OCV function.

We have thus reduced the parameter estimation problem of the linearised SPM to three independent parameters: the cathode diffusion time constant $\theta_1=\tau_+^d$, the anode diffusion time constant $\theta_4=\tau_-^d$ and the charge-transfer resistance $R_{ct}$. We will subsequently define the vector $\greekvect{\tilde{\theta}} \in \mathbb{R}^3$ of identifiable parameters as
	\begin{equation}
		\label{eq:idParameters}
		\greekvect{\tilde{\theta}} = \left[ \tau_+^d \quad \tau_-^d \quad R_{ct}^0 \right]^T .
	\end{equation}
The transfer function of the linearised SPM expressed in terms of the three parameters defined in $\greekvect{\tilde{\theta}}$ is given by
\begin{equation}
\label{eq:final_TF}
    H^0\left( s,\greekvect{\tilde{\theta}}\right)
    =
    \beta_+^0 f \left( s,\tau_+^d \right) - 
    \beta_-^0 f\left( s,\tau_-^d \right) - R_{ct}^0 .
\end{equation}

\begin{figure}
        \centering
        \includegraphics[width=0.5\textwidth]{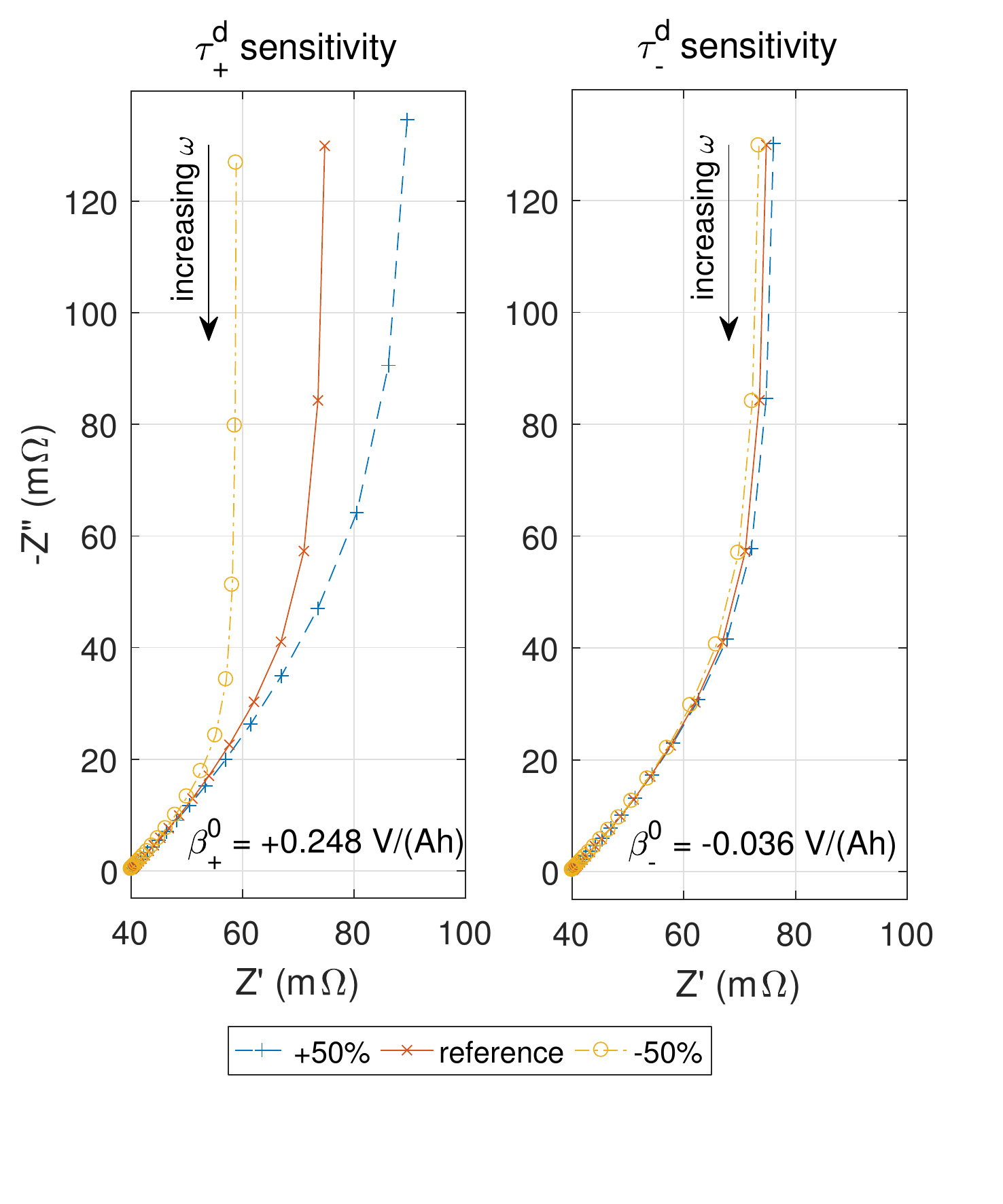}
        \caption{Comparison of the Nyquist frequency response predicted by the transfer function \eqref{eq:final_TF} using the reference parameters given in table~\ref{tab:LCO_parameters} with (left) the cathode diffusion time constant $\tau_+^d$  and (right) the anode diffusion time constant $\tau_-^d$ varied by $\pm\SI{50}{\percent}$ from their nominal values. The OCV slope values $\beta_i^0$ corresponding to the chosen linearisation point ($\text{DoD} = \SI{25}{\percent}$) are indicated on the graph for each electrode.}
        \label{fig:NyquistPlotSyntheticData}
\end{figure}

Fig.~\ref{fig:NyquistPlotSyntheticData} shows a Nyquist plot of the SPM frequency response predicted by \eqref{eq:final_TF} at \SI{25}{\percent} DoD for the LCO cell parameters given in Table~\ref{tab:LCO_parameters}. 
At low frequencies, the response features a \SI{45}{\degree} slope which is representative of semi-infinite linear diffusion. This \SI{45}{\degree} slope is commonly referred to as the `diffusion tail' in the battery electrochemical impedance spectroscopy (EIS) community and is usually modelled using a so-called Warburg constant-phase element \cite{Barsoukov2005,Buller2003a}. The cell frequency response tends towards a capacitive behaviour with decreasing frequency as exhibited by the vertical asymptote on the Nyquist plot. This capacitive behaviour arises from the variation of the average particle concentration at low frequencies, as opposed to the surface concentration variation only, due to the finite-length nature of the diffusion problem. Although the low frequencies giving rise to this capacitive behaviour are usually avoided in the lithium-ion battery EIS literature, we argue that the departure from the semi-infinite diffusion \SI{45}{\degree} tail actually provides a more informative dataset for the parameter estimation of the SPM.
Fig.~\ref{fig:NyquistPlotSyntheticData} also shows the effect of varying the diffusion time constants $\tau_+^d$ and $\tau_-^d$ in the cathode and anode respectively by $\pm \SI{50}{\percent}$ from their nominal values. Generally, a larger diffusion time constant (slower diffusion) yields higher cell impedance as expected. Moreover a smaller diffusion time constant (faster diffusion) results in the capacitive behaviour occurring at higher frequencies; this is because faster diffusion results in easier variations of the average particle concentration. Finally, the frequency response is much more sensitive to the cathode diffusion than the anode diffusion at this DoD because of the flat anode OCV curve here, as shown by the low value of the anode OCV slope $\beta_-^0$ compared to $\beta_+^0$. This behaviour will be explained in more detail in the subsequent sections.

\subsection{Structural identifiability analysis}
\label{subsec:structuralIdentifiability}
Definition~\ref{def:identifiability} can now be applied to the SPM transfer function (\ref{eq:final_TF}). To check the structural identifiability of the linearised model, we need to show that
	\begin{multline}
		\beta_+^0 f\left(s,\tau_+^d\right)  - 
		\beta_-^0 f\left(s,\tau_-^d\right) 
		- R_{ct}^0
		= \\ 		\beta_+^0 f\left(s,\tau_+^{d\ast}\right)  - 
		\beta_-^0 f\left(s,\tau_-^{d\ast}\right) 
		- R_{ct}^{0\ast}
	\end{multline}
for almost all $s$, implies the parameters equality,
	\begin{equation}
		\left[ \tau_+^d \quad \tau_-^d \quad R_{ct}^0\right]^T
		=
		\left[ \tau_+^{d\ast} \quad \tau_-^{d\ast} \quad  R_{ct}^{0\ast} \right]^T.
	\end{equation}
Because the charge transfer resistance is the only additive parameter independent of $s$ on both sides of the equation, we clearly have $R_{ct}^0 = R_{ct}^{0\ast}$, and the structural identifiability equation reduces to
	\begin{multline}
		\beta_+^0 f\left(s,\tau_+^d\right) - \beta_-^0 f\left(s,\tau_-^d\right)
		= \\
		\beta_+^0 f\left(s,\tau_+^{d\ast}\right) - \beta_-^0 f\left(s,\tau_-^{d\ast}\right),
	\end{multline}
for almost all $s$.
Since $f$ is a non-trivial function of $s$, this equality holds in the general case for most values of $s$, if and only if $\tau_+^d = \tau_+^{d\ast}$ and $\tau_-^d = \tau_i^{d\ast}$, and so the linearised SPM is structurally identifiable.
There are however a few cases where other solutions exist:
\begin{itemize}
\item If $\beta_+^0=0$ (resp.\ $\beta_-^0=0$), then any pair $\left(\tau_+^d,\tau_+^{d\ast}\right)$ (resp.\ $\left(\tau_-^d,\tau_-^{d\ast}\right)$) satisfy the identifiability equation and the linearised SPM becomes unidentifiable. This makes sense because a `flat' OCV function hides any diffusion dynamics effect of that electrode as shown in Fig.~\ref{fig:NyquistPlotSyntheticData} for the anode diffusion when $\mid\beta_-^0\mid \ll \mid\beta_+^0\mid$ .
\item If the magnitudes of the OCV function in each electrode are equal $\beta_+^0 = -\beta_-^0=\beta^0$, then interchanging the diffusion time constants $\tau_+^d=\tau_-^{d\ast}$ and $\tau_-^d=\tau_+^{d\ast}$ satisfies the structural identifiability equation, and the linearised SPM is structurally identifiable provided the diffusion time constants are ordered.
\end{itemize}
In conclusion, the linearised SPM is structurally identifiable in the general case. In practice, the fact that a `flat' OCV function results in unidentifiable parameters is not problematic since identification may be performed using data at several DoDs, ensuring the OCV functions have a significant slope in each electrode.

\section{Frequency-domain parameter estimation}
This section discusses the implementation of the estimation algorithm for the transfer function of the linearised SPM from frequency-domain EIS experimental data. The experimental methodology is discussed in section~\ref{sec:resultsAndDiscussion}. We define the vector $\greekvect{\theta}_{id}$ of parameters identified by the estimation algorithm in the general sense, where $\greekvect{\theta}_{id}$ could be equal to the full vector of parameters $\greekvect{\theta}$ given by \eqref{eq:fullParameterDef} or a subset of parameters such as $\greekvect{\tilde{\theta}}$ given by \eqref{eq:idParameters}.
The estimation is performed by finding the best fit in the least-squares sense, i.e.\ by minimising the sum of squared errors between both the real and imaginary parts of the experimental and predicted impedances over a number of frequencies \cite{Ljung1999}. First, the loss function for the estimation at a single DoD is presented and extended to simultaneous estimation at several DoDs. Then, we show how all combined purely resistive parameters, including charge-transfer resistance $R_{ct}$, can be estimated separately by simple linear regression at a given DoD.

\subsection{Single DoD parameter estimation}
The measured impedance of a cell at a given frequency $\omega_i$ is denoted by the complex number $Z_j\left( \omega_i \right) = Z_j'\left( \omega_i \right) + i Z_j'' \left( \omega_i \right)$, where $Z'$ and $Z''$ denote the real and imaginary part respectively, and the subscript $j$ denotes the DoD. Similarly, the impedance predicted by the SPM will be denoted $H_j\left( \omega_i,\greekvect{\theta}_{id} \right)$ in the subsequent sections. 

The optimal parameter estimate in the least-squares (maximum likelihood) sense is given by the estimator $\greekvect{\hat{\theta}}_{id}$ which minimises the loss function $L_{j}\left( \greekvect{\theta}_{id}\right)$ according to
% 	\begin{equation}
% 	\label{eq:argminLossFunctionSingleDoD}
% 		\greekvect{\hat{\theta}}_{id} = 
% 		\argmin_{\greekvect{\theta}_{id}} L_j \left(\greekvect{\theta}_{id}\right) =
% 		\sum_{i=1}^{N_{\omega}}
% 		\left| Z_j \left(\omega_i\right) - H_j \left(\omega_i,\greekvect{\theta}_{id}\right) \right|^2.
%      	\end{equation}
	\begin{equation}
	\label{eq:argminLossFunctionSingleDoD}
		\greekvect{\hat{\theta}}_{id} = 
		\argmin_{\greekvect{\theta}_{id}} L_j \left(\greekvect{\theta}_{id}\right) ,
	\end{equation}
	where the loss function is the sum over $N_{\omega}$ chosen frequencies of the squared magnitude of the error between predicted and measured complex impedance according to
	\begin{equation}
	L_j \left(\greekvect{\theta}_{id}\right) =
		\sum_{i=1}^{N_{\omega}}
		\left| Z_j \left(\omega_i\right) - H_j \left(\omega_i,\greekvect{\theta}_{id}\right) \right|^2.
     \end{equation}
% This loss function is the sum over $N_{\omega}$ chosen frequencies of the squared magnitude of the error between predicted and measured complex impedance. 
This is also equal to the sum of the squared real and imaginary part errors according to
	\begin{eqnarray}
		\label{eq:lossFunctionSingleDoD}
		L_j \left(\greekvect{\theta}_{id}\right) {}={}&
		\sum_{i=1}^{N_{\omega}} 
		\left( Z'_j(\omega_i) - H'_j \left( \omega_i,\greekvect{\theta}_{id} \right) \right)^2\nonumber\\
		&{+}\:\sum_{i=1}^{N_{\omega}}
		\left( Z''_j \left(\omega_i\right) - H''_j \left(\omega_i,\greekvect{\theta}_{id}\right)\right)^2.
	\end{eqnarray}

\subsection{Combined DoDs parameter estimation}
As mentioned in section~\ref{subsec:structuralIdentifiability} and shown in section~\ref{sec:resultsAndDiscussion}, the estimation of the parameter $\greekvect{\tilde{\theta}}$ based on the impedance data at a single DoD may be unidentifiable in practice due to the flatness of one of the electrode OCV functions.	
Therefore, the parameter estimation must be performed against impedance measured at several DoDs.
As before, the optimal parameter estimate is given by the argument $\greekvect{\hat{\theta}}_{id}$ minimising the loss function according to
	\begin{equation}
		\greekvect{\hat{\theta}}_{id} = \argmin_{\greekvect{\theta}_{id}} L \left( \greekvect{\theta}_{id} \right),
	\end{equation}
where the loss function $L\left( \greekvect{\theta}_{id} \right)$ in this case is simply defined as the sum of the loss functions $L_{j}$ defined in~\eqref{eq:lossFunctionSingleDoD} over $N_{DoD}$ levels of DoD, where $j$ denotes the DoD, according to:
	
    \begin{equation}
		L \left( \greekvect{\theta}_{id} \right)= \sum_{j=1}^{N_{DoD}}\sum_{i=1}^{N_{\omega}}
		\left| Z_j \left(\omega_i\right) - H_j \left(\omega_i,\greekvect{\theta}_{id} \right) \right|^2.
		\label{eq:lossFunctionCombinedDoD}
	\end{equation}

\subsection{Estimation of purely resistive terms including charge-transfer resistance}
\label{subsec:linearRegression}
Although the simultaneous parameter estimation at several DoDs could in theory improve the identifiability of the estimation problem (assuming the diffusion parameters are not a function of DoD), it also adds $N_{DoD}$ charge transfer resistances $R_{ct}^0$ to identify because of the dependency of $R_{ct}^0$ on stoichiometry (or equivalently DoD) in \eqref{eq:chargeTransferResistance}. The charge-transfer resistance $R_{ct}^0$ is an additive purely resistive term in the transfer function describing the cell impedance.
Therefore, its only effect is to shift the impedance response along the real-axis on the Nyquist plot. 
In order to reduce the parameter space to be explored, a purely resistive term denoted $R_0$, combining the higher-frequency charge-transfer resistance $R_{ct}^0$, the ohmic contact resistance of the cell and any resistive effect due to passivation layers, can be estimated separately by simple linear regression and subtracted from the experimental frequency-response data.
Recalling from section~\ref{subsec:linearisedTF} that the semi-infinite diffusion-driven impedance at lower frequencies features a characteristic \SI{45}{\degree} slope, the charge-transfer resistance can be estimated by fitting a \si{45}{\degree} straight line to the experimental response at lower frequencies, and extrapolating to find the intercept of this with the real axis on the Nyquist diagram. Although the SPM could be extended to account for double-layer capacitance effects, this is beyond the scope of the present study. Therefore, the semi-circle at higher frequencies on the experimental EIS Nyquist plot, which arises from the parallel combination of charge-transfer resistance and double-layer capacitance, is ignored. Any high-frequency effects are considered as purely resistive and accounted for in the term $R_0$ estimated by the linear regression described below. We assume that any experimental high-frequency data points where charge-transfer processes are dominant (charge-transfer semicircle), have been removed; only data points at frequencies where diffusion processes are dominant remain. The linear regression for estimating the purely resistive term is illustrated in Fig.~\ref{fig:EisChargeTransferFit} for EIS data at \SI{10}{\percent} DoD.

Denoting $x_i = Z'\left(\omega_i\right)$ and $y_i = -Z''\left(\omega_i\right)$, the real part and negative imaginary part respectively of the experimental impedance at frequency $\omega_i$, the linear regression problem is
	\begin{equation}
		y_i = \beta_1 x_i + \beta_0.
	\end{equation}
The slope is set to \SI{45}{\degree} to represent the diffusion behaviour as discussed previously. Therefore $\beta_1=1$ is known and the regression problem consists of estimating the real intercept $\beta_0$ ($=-R_0$) only. 
Defining the variable $\bar{y}_i = y_i - \beta_1 x_i$, the regression problem can be written:
	\begin{equation}
		\bar{y}_i = \beta_0.
	\end{equation}
This therefore constitutes an over-determined system of $N_\omega$ equations with the only unknown $\beta_0$, which can be written
	\begin{equation}
		\bar{\vect{Y}} = \beta_0 \vect{1},
	\end{equation}
where $\bar{\vect{Y}} = \left[ \bar{y}_1 \! , \ldots \! , \bar{y}_{N_{\omega}} \right]^T \in \mathbb{R}^{N_\omega}$ is the vector containing the values of $\bar{y}_i$ at the $N_\omega$  frequencies, and $\vect{1} \in \mathbb{R}^{N_\omega}$ is a vector of all ones of length $N_{\omega}$.
The value of $\beta_0$ can easily be determined in the least-squares sense by computing
	\begin{equation}
		\beta_0 = \vect{1}^\dagger \bar{\vect{Y}} = \frac{\vect{1}^T}{N_\omega} \bar{\vect{Y}}
	\end{equation}
where the superscript $\vect{1}^\dagger$ denotes the pseudo-inverse of the vector $\vect{1}$ which is simply equal to a row vector with all components equal to $1/N_\omega$.
This linear regression must be performed using only the experimental data points in the frequency range where low-frequency pseudo-capacitive effects arising from OCV variations are negligible.
Therefore, very low frequency impedance data points must be discarded in the experimental dataset to perform a meaningful regression.
In practice this was achieved by performing the regression first on all data points, and then on an iteratively reduced dataset where the lowest frequency impedance data point was removed until the goodness-of-fit was satisfactory.
The coefficient of determination $R^2$ was used as an indicator to assess the goodness-of-fit and a threshold value of $0.98$ was chosen to stop the iterations.

\section{Results and discussion}
\label{sec:resultsAndDiscussion}

\begin{table*}[!t]
	\caption{Parameters used for generating synthetic electrochemical impedance data for an LCO cell reported in~\cite{Bizeray2015}.}
	\label{tab:LCO_parameters}
	\centering
	\vspace{10pt}
		\begin{tabular}{l l c c}

            \hline		
			Parameter & Units & Anode & Cathode \\
			\hline
			Electrode thickness $\delta_i$ 								& \SI{}{\micro\meter}						& \num{73.5} 			& \num{70.0} 			\\
			Particle radius $R_i$ 											& \SI{}{\micro\meter}						& \num{12.5} 			& \num{8.5} 				\\
			Active material volume fraction $\epsilon_i$ 			& - 												& \num{0.4382}			& \num{0.3000}			\\
			Li diffusivity in active material $D_i$ 						& \SI{}{m^2.s^{-1}} 						& \num{5.5e-14} 		& \num{1.0e-11} 		\\
			Reaction rate constant $k_i$ 								& \SI{}{m^{2.5}.mol^{-0.5}.s^{-1}}	& \num{1.764e-11}	& \num{6.667e-11}	\\
			Max. active material concentration $c_i^{max}$		& \SI{}{mol.m^{-3}} 						& \num{30555} 			& \num{51555} 			\\
			Electrode surface concentration $\mathcal{A}$ 		& \SI{}{cm^{-2}} 							& \num{982} 				& \num{982} 				\\
			Electrolyte concentration $c_e$ 							& \SI{}{mol.m^{-3}}						& \multicolumn{2}{c}{\num{1000}} 				\\
			\hline
		\end{tabular}
\end{table*}

\subsection{Parameter estimation using synthetic data}
The parameter estimation algorithm was first tested against synthetic data generated using the linearised SPM with a known set of parameters for an LCO cell from the literature~\cite{Bizeray2015}, which are summarised in Table~\ref{tab:LCO_parameters}.

\begin{figure*}[!t]
		\centering
 		\includegraphics[width=0.9\textwidth]{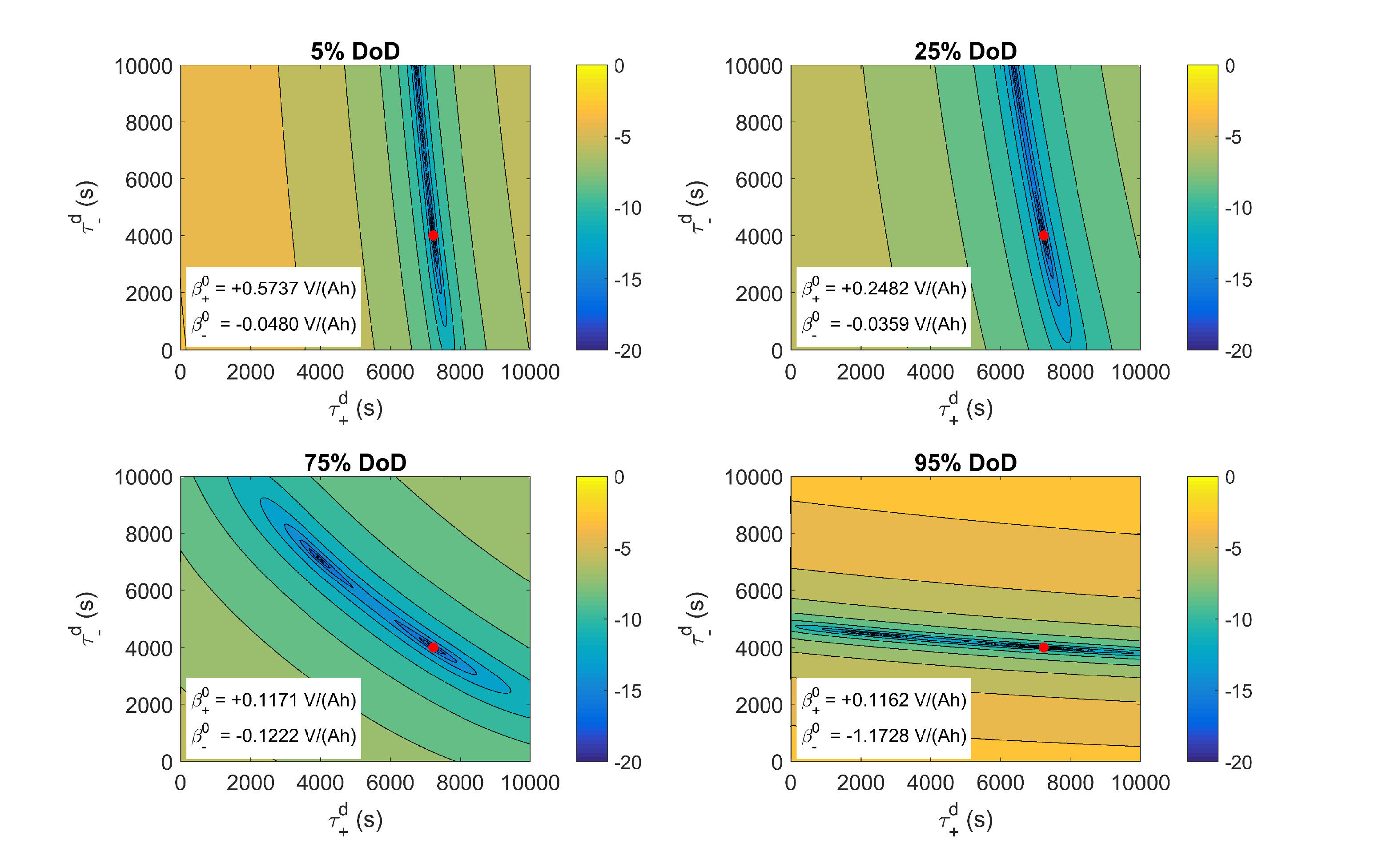}
		\caption{Contour plots of $\ln L_j \left( \greekvect{\tilde{\theta}} \right)$ against $(\tau_+^d,\tau_-^d)$ at several DoDs assuming the charge-transfer resistance $R_{ct}$ is known. Synthetic EIS data were generated using the linearised SPM with reference parameters found in the literature \cite{Bizeray2015} for an LCO cell (shown as a red dot on the contour plot).}
		\label{fig:contourPlotLnCost}		
\end{figure*}

Fig.~\ref{fig:contourPlotLnCost} shows contour plots of the natural logarithm of the loss function $L_j \left( \greekvect{\tilde{\theta}}\right)$,  \eqref{eq:lossFunctionSingleDoD}, at several DoDs for the synthetic data. It is assumed that the charge-transfer resistance $R_{ct}$ is known for each DoD since, as previously explained, it can be identified separately by straightforward linear regression. Therefore the two parameters to be identified are the cathode and anode diffusion time constants $\tau_+^d$ and $\tau_-^d$. 
The coordinates of the minimum values of the natural logarithm of \eqref{eq:lossFunctionSingleDoD} on these contour plots represent the optimal parameter estimates in the least-squares sense as defined in~\eqref{eq:argminLossFunctionSingleDoD}.
The two cases for which the linearised SPM is structurally unidentifiable, as discussed in section~\ref{subsec:structuralIdentifiability}, can be witnessed by examining Fig.~\ref{fig:contourPlotLnCost}.
Firstly, when the absolute value of the OCV functions are approximately equal in the cathode and anode, $\beta_+^0 \approx -\beta_-^0 $, such as \SI{75}{\percent}~DoD, the loss function clearly shows two minima: one minimum corresponds to the actual parameters, while the other minimum happens when the anode and cathode diffusion time constants are interchanged.
Moreover, even if the absolute values of the electrode OCV slopes $\beta_-^0$ and $\beta_+^0$ are distinct (e.g. \SIlist{5;25;95}{\percent} on Fig.~\ref{fig:contourPlotLnCost}), the loss function still features two local minima, although barely visible due to the narrow loss function along one axis. Therefore, an ambiguity persists and the anode and cathode dynamics cannot be distinguished. A possible explanation for this ambiguity is that the measured voltage is the difference between the anode and cathode OCV and not the absolute voltage of the electrode with respect to a reference electrode.
Secondly, in the case where one OCV slope is an order of magnitude lower in one electrode compared to the other, the diffusion time constant of this electrode becomes unidentifiable, e.g. \SIlist{5;25;95}{\percent}~DoD in Fig.~\ref{fig:contourPlotLnCost}. For instance at \SI{95}{\percent}~DoD, the absolute value of the cathode OCV slope $\beta_+^0$ is small compared to the anode slope, and the minimum of the loss function is elongated along the $\tau_+^d$ axis, i.e.\ provided the value of $\tau_-^d$ is correct, almost any value for $\tau_+^d$ will result in a minimum value of the loss function.

\begin{figure}
		\centering
 		\includegraphics[width=0.5\textwidth]{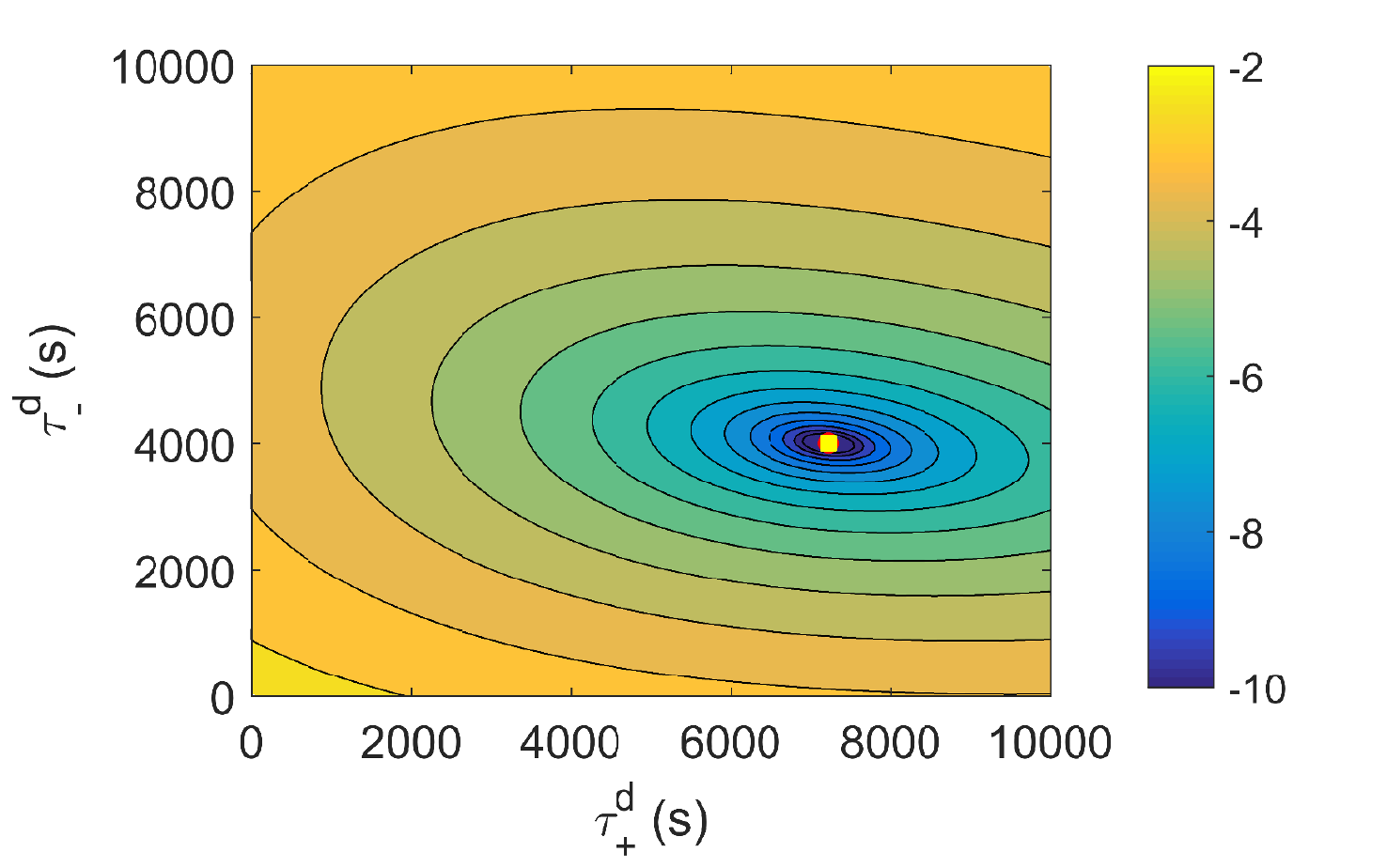}
		\caption{Contour plots of $\ln L \left( \greekvect{\tilde{\theta}} \right)$ against $(\tau_+^d,\tau_-^d)$ assuming the charge-transfer resistance $R_{ct}$ is known, based on synthetic EIS data with the reference parameters from \cite{Bizeray2015} for an LCO cell (shown as a yellow square on the contour plot). The loss function is the sum of the cost function $L_j$ for four levels of DoD \SIlist{5;25;75;95}{\percent}.}
		\label{fig:contourSumPlotLnCost}		
\end{figure}

These results suggest that the anode and cathode diffusion time constants cannot both be identified from frequency-data at a single DoD. In the best case, where the OCV slopes are equal for the chosen DoD, the values $\tau_+^d$ and $\tau_-^d$ can be determined but cannot be assigned to a specific electrode.
Fig.~\ref{fig:contourSumPlotLnCost} shows the natural logarithm of the loss function $L\left(\greekvect{\tilde{\theta}}\right)$ combining the EIS data at \SIlist{5;25;75;95}{\percent} DoD as defined in (\ref{eq:lossFunctionCombinedDoD}).
This loss function shows a single global minimum corresponding to the actual set of parameters used to generate the synthetic data.
The two time constants of the linearised SPM can therefore only be estimated unambiguously by considering EIS data at several DoDs.
More specifically, complementary DoD values, where each electrode in turn presents a large OCV slope while the other is negligible, should be combined to ensure better identifiability. For instance, it is clear from Fig.~\ref{fig:contourPlotLnCost} that combining EIS data at \SI{5}{\percent} and \SI{95}{\percent} DoD will result in a single minimum at the intersection of the lines of minimum loss function values for individual DoDs.

\subsection{Parameter estimation using experimental data}
This section discusses the results of the parameter estimation using  EIS experimental data  measured from a commercial Kokam NMC cell. First, the experimental procedure used to measure the cell impedance and the OCV for both electrodes is discussed. 
The estimation of the charge-transfer resistance using linear regression is then briefly considered. 
Finally, the practical identifiability of the model diffusion time constants from the experimental impedance data is examined.

\subsubsection{Experimental setup}
\label{subsubsec:experimentalSetup}
Electrochemical impedance spectroscopy (EIS) \cite{Barsoukov2005} is a widely used technique to measure the frequency response of electrochemical systems, including lithium-ion batteries. During a galvanostatic EIS (GEIS) experiment, the battery is excited with a small sinusoidal current and the associated voltage response is recorded. This is repeated at a number of frequencies. Potentiostatic EIS is a similar technique, where the voltage is used as the input sinusoidal signal and the recorded output is the current. The impedance of the cell at each frequency is calculated from the gain and phase shift between the input and output signals. EIS experimental data were obtained for a Kokam SLPB533459H4 \num{740}~mAh NMC cell at several DoDs using a BioLogic SP-150 potentiostat.
The cell was kept at a constant \SI{20}{\degreeCelsius} temperature in a V\"otsch~VT4002 thermal chamber in order to minimise the possible detrimental effects of temperature variations on the results of the parameter identification approach.
The cell impedance was measured using single-sine GEIS, averaged over two periods, with a peak-to-peak AC current amplitude of \SI{100}{\milli\ampere}, and \SI{0}{\milli\ampere} DC current.  The impedance was measured at logarithmically-spaced frequencies with six frequencies per decade ranging from \SI{5}{\kilo\hertz} to \SI{200}{\micro\hertz}.
A resting time of 3~hours was applied before any EIS measurement to ensure that the cell was close to equilibrium.
The EIS measurement using this experimental setup took approximately 9~hours for each DoD because of the low frequencies required. Clearly this is quite time consuming and ultimately one should aim at using the minimum number of low-frequency data and/or DoD linearisation points to accelerate the model parametrisation.
\begin{figure}
        \centering    
        \includegraphics[width=0.5\textwidth]{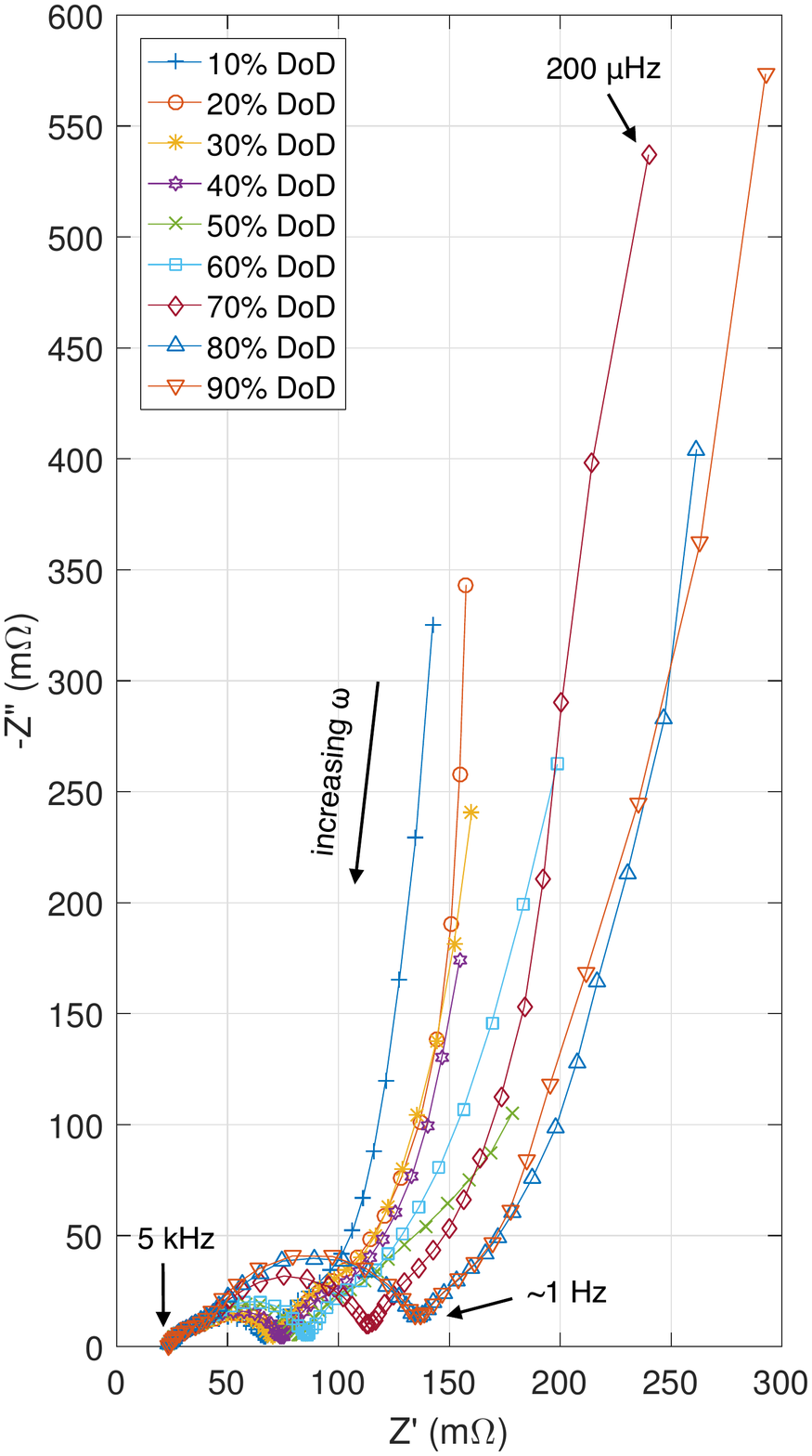}
        \caption{Nyquist plot of the electrochemical impedance of a Kokam SLPB 533459H4 \num{740}~mAh NMC cell measured at \SI{20}{\degreeCelsius} and several DoDs in the frequency range \SI{5}{\kilo\hertz} to \SI{200}{\micro\hertz}.
        \label{fig:experimentalEISNyquist}
        }
\end{figure}

Fig.~\ref{fig:experimentalEISNyquist} shows the measured EIS data as a Nyquist plot for several DoDs ranging from \SI{10}{\percent} to \SI{90}{\percent}. These impedance results are typical of lithium-ion batteries, which have smaller impedance at high frequencies (bottom left), arising from ohmic resistance and fast charge-transfer processes, and larger impedance at lower frequencies (top right), arising from slower diffusion processes and SoC variations.
At high frequencies, the response features a depressed semicircle, which is characteristic of charge-transfer processes. As discussed in section \ref{subsec:linearRegression}, this part of the plot is discarded because charge-transfer processes are approximated by the resistor $R_{ct}^0$ only in \eqref{eq:final_TF}.
As expected, in the medium range of frequencies where semi-infinite diffusion is predominant, the frequency response shows the familiar \SI{45}{\degree} diffusion slope. At very low frequencies the response tends towards a capacitive vertical line due to the variations of the average particle stoichiometry.

Fig.~\ref{fig:experimentalOCV}d shows the OCV of the Kokam SLPB533459H4 cell as a function of discharge capacity $Q$ and DoD measured experimentally. In order to measure the anode and cathode potential separately, a minimally invasive reference electrode consisting of a lithium-coated copper wire was inserted into the commercial pouch cell using the technique presented in \cite{McTurk2015}. The OCV of both electrodes were then measured using the galvanostatic intermittent titration technique (GITT) \cite{Birkl2016} by discharging the cell in \SI{14.8}{\milli\ampere\hour} increments at \textit{C}/10 and measuring the cell OCV after a one-hour voltage relaxation time to obtain 50~measurements points.

\subsubsection{Estimation of high-frequency purely resistive terms}

\begin{figure}
		\centering
 		\includegraphics[width=0.45\textwidth]{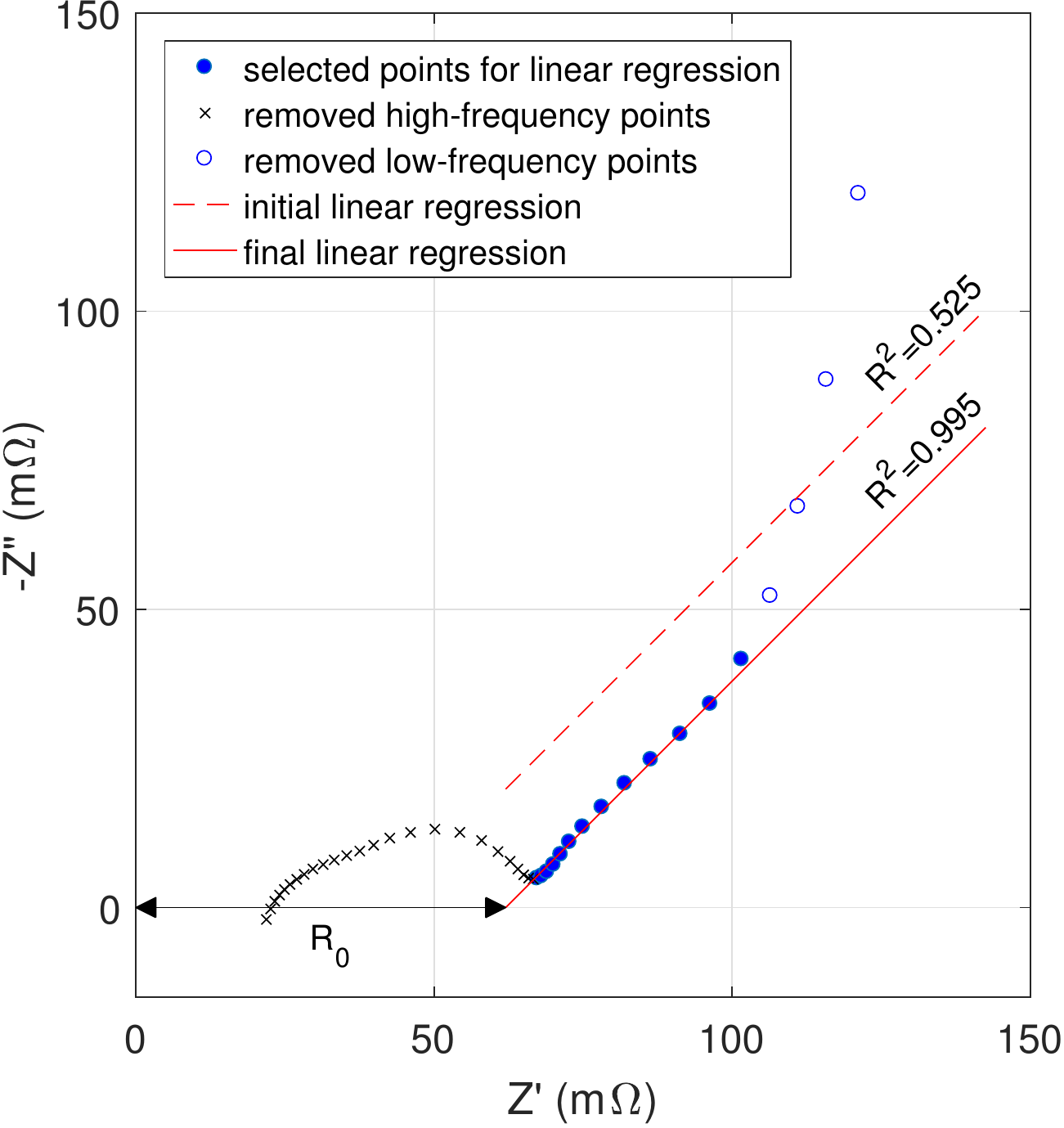}
		\caption{Linear regression of the purely resistive term $R_0$ from experimental EIS data at \SI{10}{\percent} DoD. The high-frequency data points (crosses) are discarded. The first linear regression (dashed line) is performed on all data points (circles). Low-frequency data points (empty circles), corresponding to the pseudo-capacitive effect arising from DoD variations, are iteratively removed until the final remaining set of data points (filled circles) enables a linear regression with a coefficient of determination $R^2$ greater than \num{0.98} (the three lowest frequency data points are not shown on this graph for readability).}
		\label{fig:EisChargeTransferFit}		
\end{figure}

\begin{figure}
		\centering    
 		\includegraphics[width=0.5\textwidth]{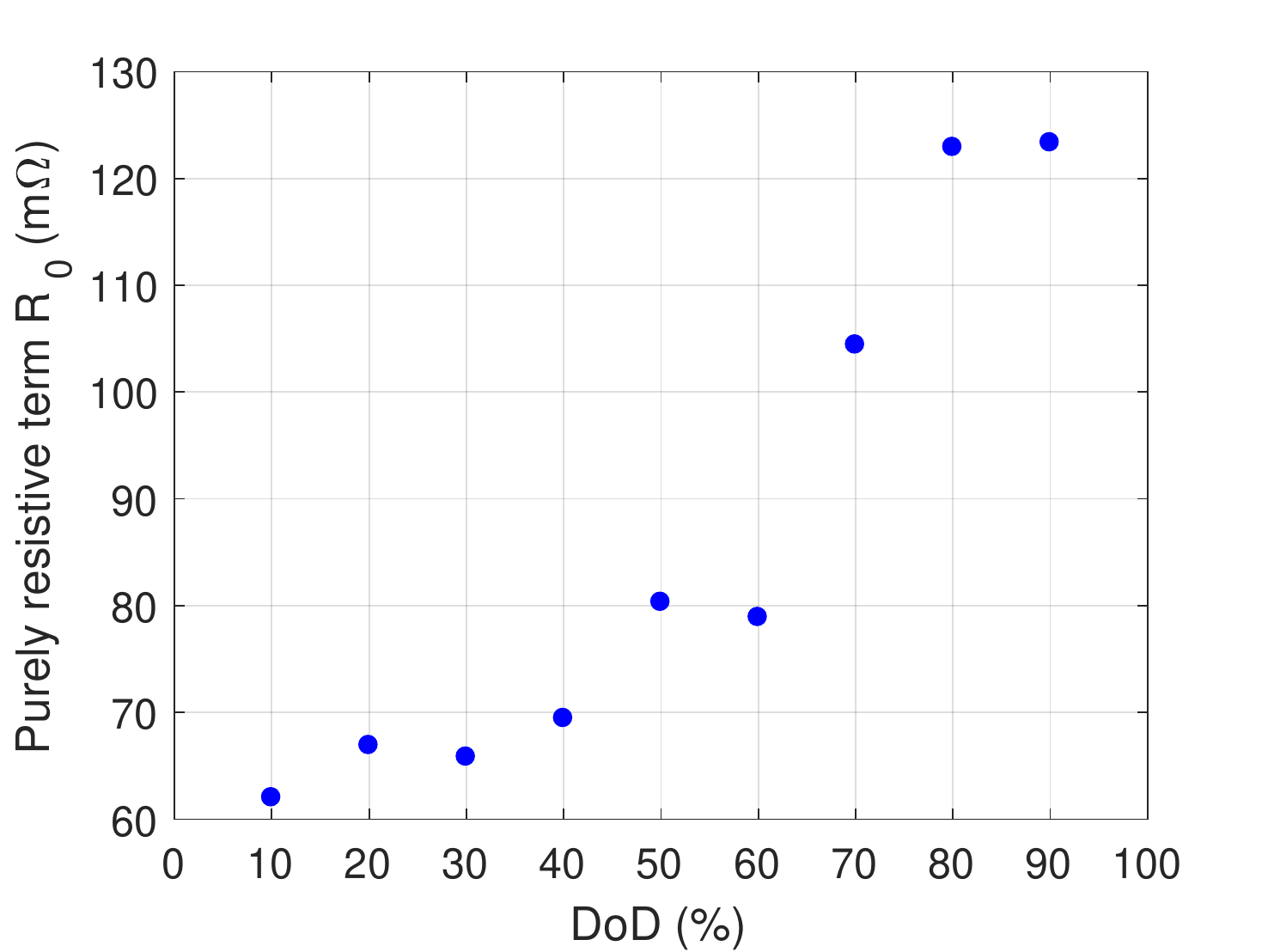}
		\caption{Purely resistive term $R_0$, accounting for charge-transfer, passivation layer and contact resistance at high frequencies,against DoD estimated from experimental data by linear regression.}
		\label{fig:chargeTransferResistanceValues}
\end{figure}

The purely resistive term $R_0$ accounting for high-frequency effects, including charge-transfer, passivation layer and contact resistance, which shifts the diffusion-driven impedance to the right on the experimental EIS Nyquist plot, is estimated using the linear regression method discussed in section~\ref{subsec:linearRegression}. 
Fig.~\ref{fig:EisChargeTransferFit} shows the results of the linear regression for the EIS data collected at \SI{10}{\percent} DoD. As discussed previously, high frequency data points corresponding to charge-transfer resistance and double-layer capacitance are discarded and only data points at frequencies where diffusion effects are predominant are considered. The initial linear regression is performed on all low frequency data points but yields very poor fitting ($R^2=0.525$) due to the the pseudo-capacitive behaviour at very low frequencies arising from DoD variations. By iteratively removing the lower frequency data points, a good linear fit of the \SI{45}{\degree} tail can be achieved ($R^2=0.995$). The purely resistive term $R_0$ is then approximated by finding the intercept of the linear regession curve with the real axis.
The estimated values of $R_0$ as a function of DoD are reported on Fig.~\ref{fig:chargeTransferResistanceValues}. Note that the cell ohmic resistance included in $R_0$ is approximately equal to \SI{25}{\milli\ohm} according to the intercept of the experimentally measured impedance with the real-axis of the Nyquist plot at high frequencies, Fig.~\ref{fig:experimentalEISNyquist}. As shown, the resistive term $R_0$ including charge-transfer and contact resistance increases with DoD, which is consistent with previously reported Kokam NMC cell EIS data in the literature \cite{Howey2014}.

\subsubsection{Parameter estimation performance}

\begin{figure*}[!t]
		\centering    
 		\includegraphics[width=0.9\textwidth]{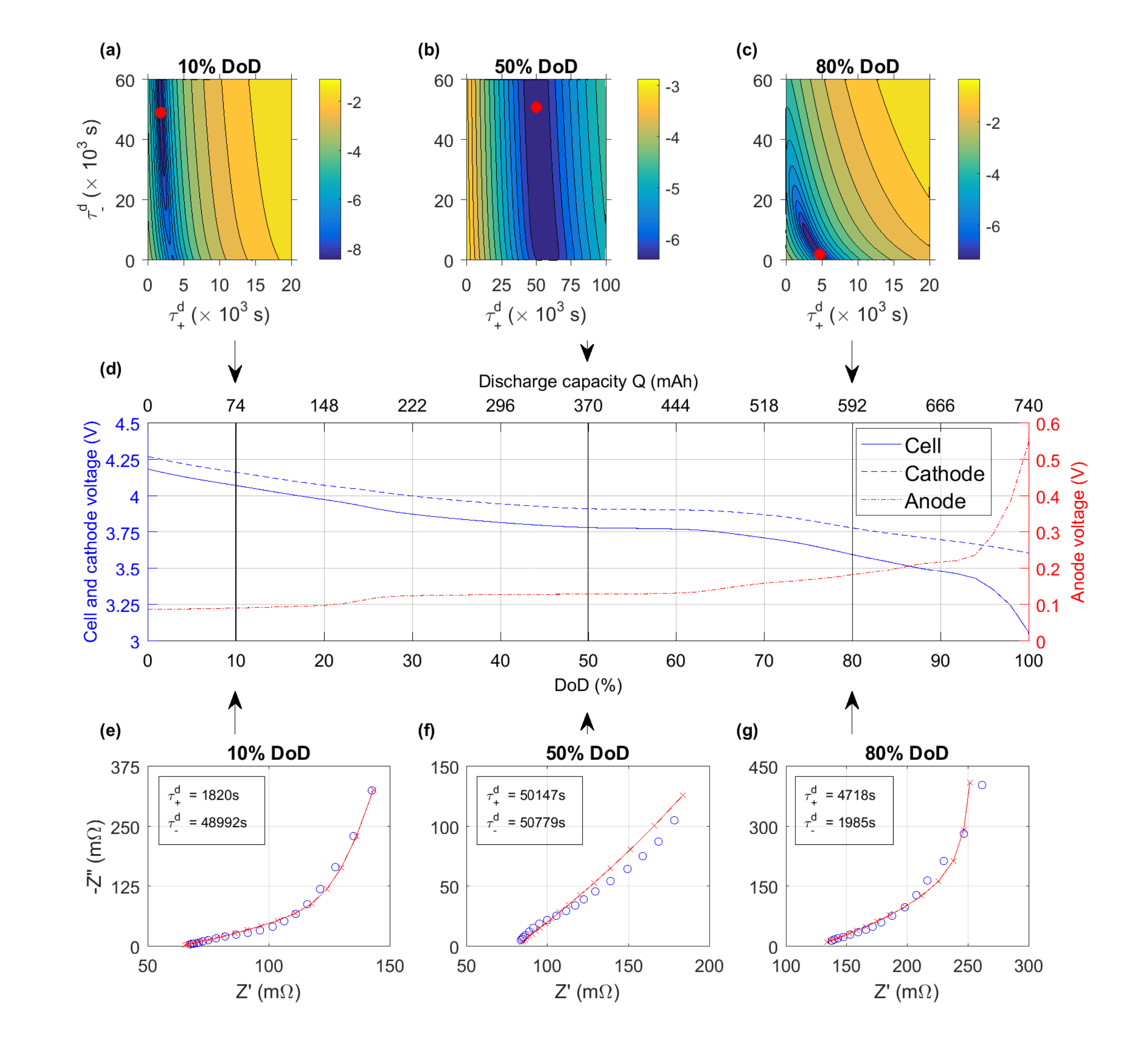}
		\caption{Natural logarithm of the individual DoD loss function $L_j \left( \greekvect{\tilde{\theta}}\right)$ based on experimental EIS data at (a) \SI{10}{\percent}, (b) \SI{50}{\percent}, and (c) \SI{80}{\percent} DoD. (d) Cathode, anode and cell OCV of the Kokam cell as a function of DoD measured by the GITT technique using a minimally invasive reference electrode. Cell impedance experimentally measured (circles) and predicted (solid-lines and crosses) by the linearised SPM transfer function model fitted by minimising the individual DoD loss function $L_j \left( \greekvect{\tilde{\theta}}\right)$ at (e) \SI{10}{\percent}, (f) \SI{50}{\percent}, and (g) \SI{80}{\percent} DoD.}
		\label{fig:experimentalOCV}
\end{figure*}

\begin{table}[!t]    
\renewcommand{\arraystretch}{1.3}
\caption{RMS and maximum error between the measured and predicted impedance at \SI{10}{\percent}, \SI{50}{\percent} and \SI{80}{\percent} DoD.}	
\label{tab:rmsAndMaxError}	
\centering	
\vspace{10pt}		
\begin{tabular}{l | c c | c c}
	  & \multicolumn{2}{c}{Single DoD estimation } & \multicolumn{2}{| c}{Combined DoDs estimation} \\ 
\hline  
DoD & RMS Error & Max Error & RMS Error & Max Error \\  
\hline 
\SI{10}{\percent} & \SI{3.28}{\milli\ohm} & \SI{5.60}{\milli\ohm} & \SI{6.66}{\milli\ohm} & \SI{16.73}{\milli\ohm}  \\
\SI{50}{\percent} & \SI{9.56}{\milli\ohm} & \SI{21.36}{\milli\ohm} & \SI{47.88}{\milli\ohm} & \SI{95.16}{\milli\ohm} \\
\SI{80}{\percent} & \SI{6.48}{\milli\ohm} & \SI{11.32}{\milli\ohm} & \SI{6.84}{\milli\ohm} & \SI{12.82}{\milli\ohm} \\ 
\hline		
\end{tabular}\end{table}

Fig.~\ref{fig:experimentalOCV} presents practical identifiability and parameter estimation results based on experimental impedance data at individual DoDs. Three characteristic values of DoD where chosen, \SIlist{10;50;80}{\percent}, in order to demonstrate the identifiability issues discussed in section~\ref{subsec:structuralIdentifiability}.
At \SI{10}{\percent}~DoD, only the cathode OCV shows a significant slope, at \SI{50}{\percent} DoD both the cathode and anode OCV are flat, and at \SI{80}{\percent} DoD both electrodes feature a significant OCV slope. The experimental and fitted impedance responses are shown on Fig.~\ref{fig:experimentalOCV}e, f and g for the chosen values of DoD. The corresponding root-mean square (RMS) and maximum (Max) error between the predicted and measured impedance response for this single DoD parameter estimation algorithm are presented in Table~\ref{tab:rmsAndMaxError}. A satisfactory fit is obtained in all cases with a root-mean square error of \SIlist{3.28;9.56;6.48}{\milli\ohm} for \SIlist{10;50;80}{\percent} DoD respectively.

However, the uncertainties on the parameter estimates are large as exhibited by the contour plots of the loss function in Fig.~\ref{fig:experimentalOCV}a, b and c.
At \SI{10}{\percent} DoD the loss function is very sensitive to the cathode diffusion dynamics while anode dynamics have very little impact because of the very small slope of the anode OCV. Therefore the estimated anode diffusion time constant $\tau_-^d$ is highly uncertain and cannot be estimated from this data set alone.
In contrast, at \SI{80}{\percent} DoD the loss function is sensitive to both the anode and cathode dynamics because of the similar magnitude of the OCV slope in both electrodes. However, the minimum of the loss function is still relatively elongated along the anode parameter axis, which suggests that the uncertainty on the anode parameter remains larger than the cathode one. This confirms that impedance data at several DoDs must be combined to yield less uncertain parameter estimates.

At \SI{50}{\percent} DoD, where the OCV is flat for both electrodes, the loss function is completely insensitive to the anode dynamics and somewhat more sensitive to the cathode dynamics. However, the contour plot of the loss function is actually shallow along the cathode parameter as well, implying that the uncertainty on both parameters is very large.
Furthermore, the minimum of the loss function happens for a cathode diffusion time constant centred around \SI{60000}{\second} as shown in Fig.~\ref{fig:experimentalOCV}b. This is approximately an order of magnitude higher than the value of $\tau_+^d$ estimated using the other data sets at \SIlist{10;80}{\percent} DoD.
Throughout this work, the implicit assumption was made that both diffusion time constants do not vary with DoD. However, Fig.~\ref{fig:experimentalOCV}b suggests that lithium diffusivity in the electrode active material could drastically change with DoD. 
Experimental measurements of lithium diffusivities in the active material of a Kokam NMC cell are reported in \cite{Ecker2015} and confirm this. More especially, both the anode and cathode diffusivities are an order of magnitude lower around \SI{50}{\percent} DoD compared to high DoDs \cite{Ecker2015}.
Another source of uncertainty that has been ignored in the model is the error related to the OCV measurement and the calculation of the OCV slopes $\beta_+^0$ and $\beta_-^0$. At \SI{50}{\percent} DoD, the OCV slopes in both electrodes are very small and therefore the relative error on measurement is large, especially considering that numerical differentiation is highly sensitive to noise.
Finally, because the OCV slopes are flat, both the anode and cathode diffusion dynamics have very limited impact on the cell terminal voltage at \SI{50}{\percent} DoD.
Therefore, the EIS data at \SI{50}{\percent} DoD will be discarded from the parameter estimation using combined DoDs discussed subsequently.

\begin{figure}
    \centering    
    \includegraphics[width=0.45\textwidth]{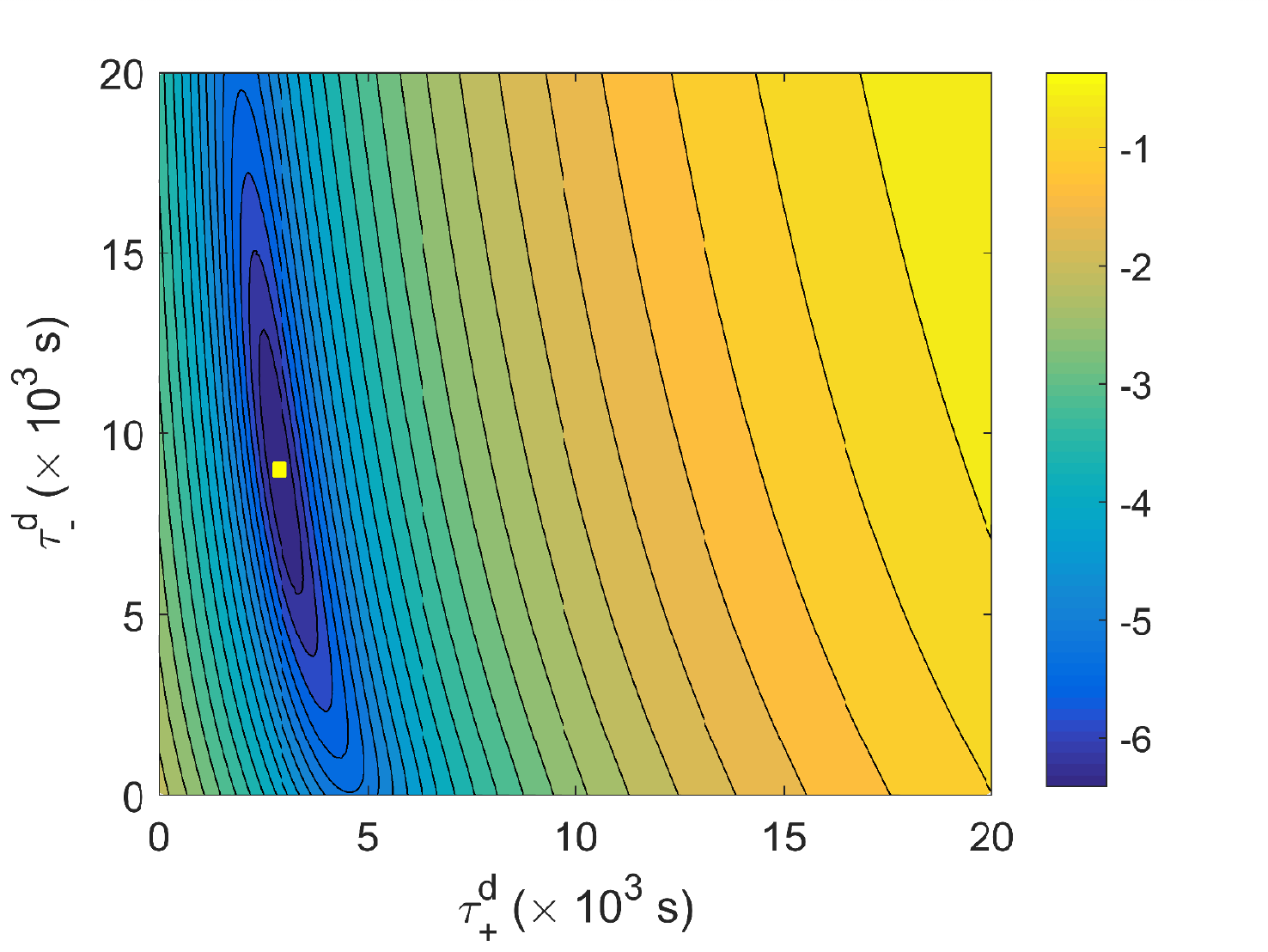}
    \caption{Contour plot of the loss function $\ln L\left(\greekvect{\tilde{\theta}}\right)$, (\ref{eq:lossFunctionCombinedDoD}), based on experimental impedance data at \SIlist{10;80}{\percent} DoD. The yellow square indicates the parameter estimated by the PE algorithm at $\tau_+^d=\SI{2880}{\second}$ and $\tau_-^d=\SI{9033}{\second}$.}
    \label{fig:experimentalCostFunctionCombinedDoD}
\end{figure}

Fig.~\ref{fig:experimentalCostFunctionCombinedDoD} shows the contour plot of the cost function $L\left(\greekvect{\tilde{\theta}}\right)$ combining the experimental impedance data measured at \SIlist{10;80}{\percent} DoDs.
Similarly to the results with synthetic data, Fig.~\ref{fig:contourSumPlotLnCost}, the loss function presents a single minimum, elongated along the anode parameter axis suggesting a larger uncertainty on the anode parameter compared to the cathode one. 
Although this suggests a `well-behaved' estimation problem with a single parameter estimate, these results must be considered with caution. Indeed, the two combined datasets at \SIlist{10;80}{\percent} DoDs used to perform the parameter estimation were chosen somewhat arbitrarily. Although these parameter estimates yield a good impedance fit for these DoDs, it results in very approximate fitted impedance at other DoDs, e.g. \SI{50}{\percent} DoD. As shown in Table~\ref{tab:rmsAndMaxError}, the impedance prediction error is much higher at \SI{50}{\percent} DoD with \SI{47.88}{\milli\ohm} RMS error compared to \SI{6.66}{\milli\ohm} and \SI{6.84}{\milli\ohm} at \SI{10}{\percent} and \SI{80}{\percent} DoDs respectively.
Firstly, as previously mentioned the OCV slopes $\beta_i^0$ in both electrodes are assumed fixed and perfectly known for each DoD. The sensitivity of the predicted impedance response to the OCV slopes is not negligible.
The computation of the OCV slope is highly sensitive to measurement noise because the measured voltage variations are relatively small between each DoD, on the order of a few millivolts.
Additionally, the measured OCV might not reflect closely enough the actual cell OCV. The OCV measurements were performed on a different cell (of the same type and size) which was modified by inserting a reference electrode. Although this technique is minimally invasive, it nonetheless may slightly modify the cell behaviour. Finally, electrode OCV is affected by temperature variation and hysteresis effects \cite{Birkl2016}, which introduce further uncertainties.
Secondly, our parameter estimation algorithm combining experimental data at several DoDs assumes that the diffusion time constants in both electrodes remain constant with DoD. Although solid-phase diffusivities are usually assumed constant with respect to active material lithiation in the lithium-ion battery electrochemical modelling community, e.g. \cite{Bizeray2015,Northrop2011b,Schmidt2010,Fuller1994,Guo2011}, this assumption may introduce additional inaccuracies on the parameter estimate, especially for NMC cells \cite{Ecker2015}. All of these challenges provide fruitful avenues for further exploration.

\subsection{Model validation in the time domain}
To demonstrate the performance of the linearised SPM using our frequency domain parameter estimation approach, we also include here a brief comparison between the modelled and measured voltage response in the time domain. The measurements were made on the same Kokam cell type as previously described, using the same equipment that was used for EIS measurements. The current demand was specified based on a dynamic loading profile at \SI{1}{\hertz} derived from an electric vehicle drive cycle, see \cite{Zhao2017}. The maximum current was \SI{4.4}{\ampere}, which corresponds to a fairly high \textit{C}-rate (almost 6\textit{C}). The same current input signal used in the experimental measurements was provided to the linearised SPM, and the terminal voltage was calculated by solving the model in the time domain. 
This requires solution of (\ref{eq:diffEq_groups_variableChange}-\ref{eq:initEq_groups_variableChange}, \ref{eq:linearisedVoltageMeas}, \ref{eq:beta_i}), which requires knowledge of $Q_i^{th}$, the theoretical capacity of each electrode, which relates variations in stoichiometry $x$ to variations in measured discharge capacity. Unfortunately $Q_i^{th}$ is unknown, but by introducing two changes of variable, $\hat{u}_i = Q_i^{th} \bar{u}_i$ and $ \hat{x}_i^s = Q_i^{th} \bar{x}_i^s$, this term can be eliminated from the linearised SPM, similarly to section \ref{subsec:linearisedTF}. This approach assumes the model is linearised; to run the SPM with non-linear OCV would require knowledge of $Q_i^{th}$.
The time-domain linearised SPM was solved in MATLAB\textsuperscript{\textregistered} using Chebyshev orthogonal collocation as in our previous work \cite{Bizeray2015, Bizeray2016}.

\begin{figure}
		\centering
 		\includegraphics[width=0.45\textwidth]{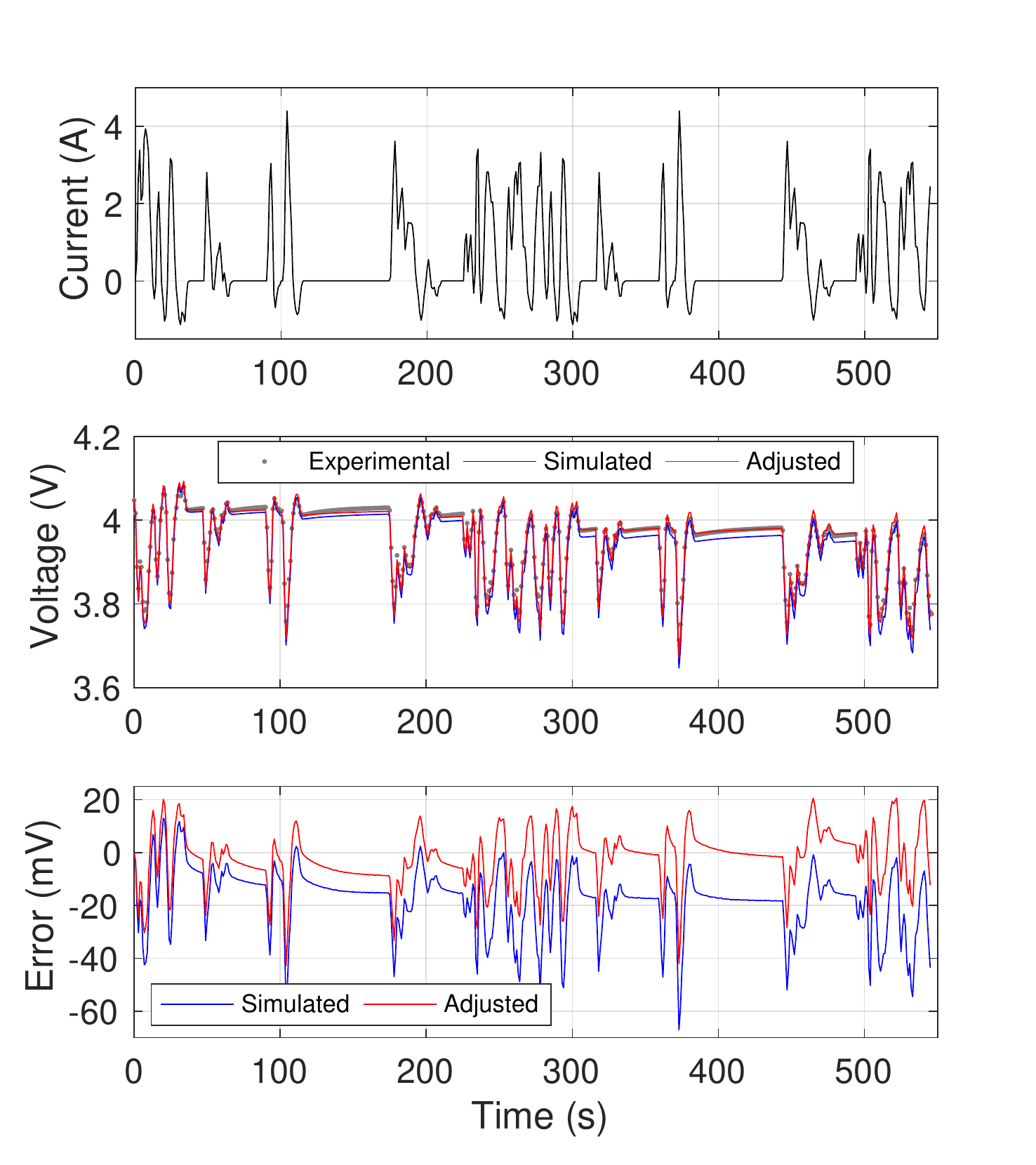}
		\caption{Time domain comparison of SPM against measured data. Start point is \SI{10}{\percent} DoD and end point is \SI{20}{\percent} DoD.}
		\label{fig:timeDomainValidation}		
\end{figure}

The comparison between linearised SPM and measured data is shown in Fig.~\ref{fig:timeDomainValidation} over a time period of almost 10 minutes. The DoD at the beginning of the period was \SI{10}{\percent} and at the end of the period was \SI{20}{\percent}. The OCV curves are almost linear over this range (see Fig.~\ref{fig:experimentalOCV}) and the variation in $R_0$ is small (Fig.~\ref{fig:chargeTransferResistanceValues}). Initial results using this approach showed a consistent increasingly negative offset in voltage predictions over the time period examined. This is most likely due to the assumed OCV slope at this DoD being inaccurate, and was adjusted by multiplying the linearised cathode OCV slope by a factor of 0.78. Fig.~\ref{fig:timeDomainValidation} shows both the initial simulated and the adjusted simulated data in comparison to the measured data. The voltage errors between adjusted simulated vs.\ measured data are maximum \SI{20}{\milli\volt} and RMS \SI{10}{\milli\volt}. These are satisfyingly reasonable given the fact the model was parametrised completely separately with frequency domain data, that different cells were used for OCV measurements compared to EIS and time domain data, and that the kinetics were linearised despite current peaks of up to 6\textit{C}.

\section{Summary and conclusion}
The analysis of parameter identifiability from experimental data is crucial prior to any attempt at estimating the parameters of battery first-principle models. We have demonstrated that the lithium-ion battery single particle model is over-parametrised and that only six subgroups of parameters are necessary for full parametrisation. Assuming that the OCV vs.\ discharge capacity curves for each electrode are known, we have shown that the estimation of the linearised model parameters at a given DoD can only identify three parameters among these six subgroups of parameters, namely the cathode diffusion time constant $\tau_+^d$, the anode diffusion time constant $\tau_-^d$ and a charge-transfer resistance $R_{ct}$.
Finally, we have shown the crucial role of the slope of electrode open-circuit voltage curves on the parameter uncertainty. This has been investigated through analysis of the model, as well as through fitting both time and frequency domain data. A flat electrode OCV curve results in high uncertainties on the electrode diffusion parameter. Therefore impedance data at a single DoD cannot, in general, result in accurate parameter estimation for both electrodes, and experimental data at multiple DoDs must be considered. Specifically, complimentary DoDs where in turn the anode and cathode OCV gradients are significant must be chosen to yield satisfactory identifiability. Future work could investigate the effects of temperature on parameter estimation, and the evolution of cell parameters as a battery ages.

\section*{Acknowledgment}
Financial support is gratefully acknowledged from EPSRC UK (EP/K002252/1) and Samsung Electronics Co. Ltd. We thank Dr~Adam Mahdi (University of Oxford) for his valuable advice on model structural identifiability analysis, and Dr Christoph Birkl and Dr Shi Zhao (University of Oxford) for providing the experimental OCV and dynamic test data.

\ifCLASSOPTIONcaptionsoff
  \newpage
\fi

\begin{IEEEbiography}[{\includegraphics[width=1in,height=1.25in,clip,keepaspectratio]{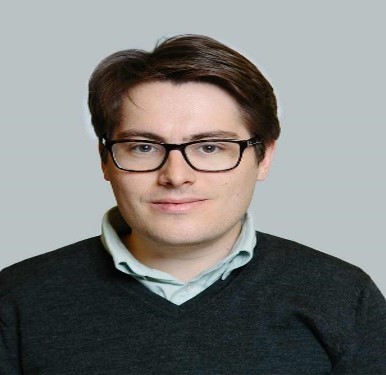}}]{Adrien M. Bizeray} received the D.Phil degree from the University of Oxford in 2017. His research interests focus on the mathematical modelling of lithium-ion batteries for developing the next-generation of Battery Management Systems relying on electrochemical model-based state estimation and control algorithms for improved battery safety and lifetime. Previously, he received the French Engineering Diploma from Arts et M\'etiers ParisTech, France, and the M.Sc. in Advanced Mechanical Engineering from Imperial College London, U.K.
\end{IEEEbiography}

% \begin{IEEEbiography}[]{Jin--Ho Kim}
% \end{IEEEbiography}

\begin{IEEEbiography}[{\includegraphics[width=1in,height=1.25in,clip,keepaspectratio]{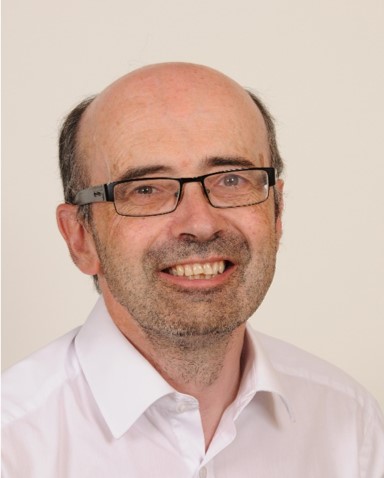}}]{Stephen R. Duncan}
is a Professor in the Department of Engineering Science at the University of Oxford, where he is a member of the Control group. He is also a Fellow of St Hugh’s College. Prior to joining Oxford, he was Reader in the Control Systems Centre at the University of Manchester (1993 to 1998) and a Director of Greycon Limited (a spin-out company from Imperial College). He has an MA in Physics from the University of Cambridge and an MSc and PhD in Control Systems from Imperial College, London. He is a member of the Editorial Board of the IET Proceedings Control Theory and Applications.
\end{IEEEbiography}

\begin{IEEEbiography}[{\includegraphics[width=1in,height=1.25in,clip,keepaspectratio]{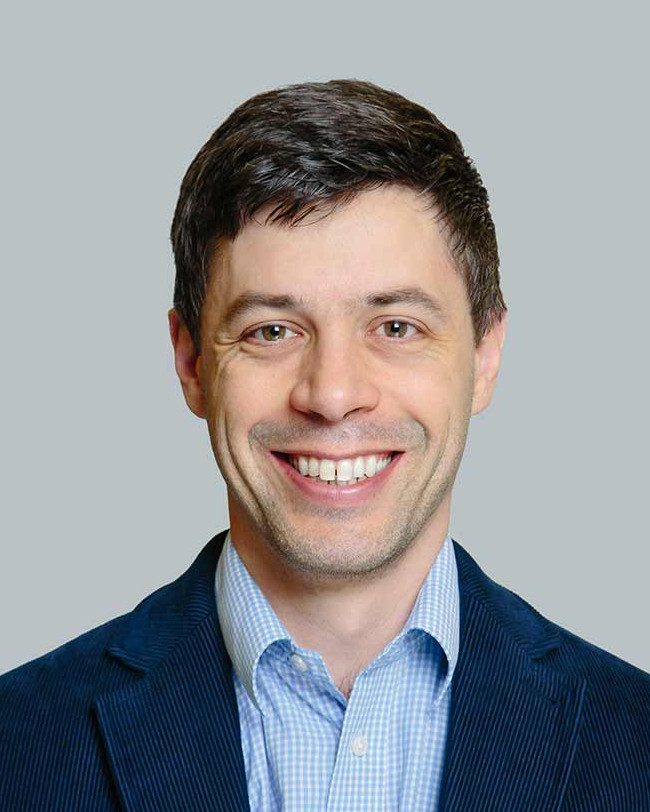}}]{David A. Howey}
(M’10) received the B.A. and
M.Eng. degrees from Cambridge University, Cambridge, U.K., in 2002, and the Ph.D. degree from the Imperial College London, London, U.K., in 2010. He is currently an Associate Professor with the Energy and Power Group, Department of Engineering Science, University of Oxford, Oxford, U.K. He was involved in projects on battery degradation diagnostics and prognostics, battery modeling, optimal control of charging and discharging, energy storage and microgrids, thermal management, novel pack designs, and motor degradation. His current research interests include condition monitoring and management of electric, and hybrid vehicle components.
\end{IEEEbiography}

% You can push biographies down or up by placing
% a \vfill before or after them. The appropriate
% use of \vfill depends on what kind of text is
% on the last page and whether or not the columns
% are being equalized.

%\vfill

% Can be used to pull up biographies so that the bottom of the last one
% is flush with the other column.
%\enlargethispage{-5in}

% that's all folks

\begin{thebibliography}{50}
\expandafter\ifx\csname natexlab\endcsname\relax\def\natexlab#1{#1}\fi
\providecommand{\url}[1]{\texttt{#1}}
\providecommand{\href}[2]{#2}
\providecommand{\path}[1]{#1}
\providecommand{\DOIprefix}{doi:}
\providecommand{\ArXivprefix}{arXiv:}
\providecommand{\URLprefix}{URL: }
\providecommand{\Pubmedprefix}{pmid:}
\providecommand{\doi}[1]{\href{http://dx.doi.org/#1}{\path{#1}}}
\providecommand{\Pubmed}[1]{\href{pmid:#1}{\path{#1}}}
\providecommand{\bibinfo}[2]{#2}
\ifx\xfnm\relax \def\xfnm[#1]{\unskip,\space#1}\fi
%Type = Article


\bibitem{Bizeray2015}
\bibinfo{author}{A.~M. Bizeray}, \bibinfo{author}{S. Zhao}, \bibinfo{author}{S.~R. Duncan}, \bibinfo{author}{D.~A. Howey},
\newblock \bibinfo{title}{Lithium-ion battery thermal-electrochemical model-based state estimation using orthogonal collocation and a modified extended Kalman filter}.
\newblock \bibinfo{journal}{J. Power Sources}, \bibinfo{volume}{296}, \bibinfo{pages}{400--412} (\bibinfo{year}{2015})

\bibitem{Klein2010}
\bibinfo{author}{R. Klein}, \bibinfo{author}{N.~A. Chaturvedi}, \bibinfo{author}{J. Christensen}, \bibinfo{author}{J. Ahmed}, \bibinfo{author}{R. Findeisen}, \bibinfo{author}{A. Kojic},
\newblock \bibinfo{title}{State estimation of a reduced electrochemical model of a lithium-ion battery}.
\newblock \bibinfo{journal}{Proc. Am. Control Conf. ACC~2010},
\bibinfo{city}{Baltimore, MD, USA},
 \bibinfo{pages}{6618--6623} (\bibinfo{year}{2010})

\bibitem{Moura2013a}
\bibinfo{author}{S.~J. Moura}, \bibinfo{author}{N.~A. Chaturvedi}, \bibinfo{author}{M. Krstic},
\newblock \bibinfo{title}{Constraint Management in Li-ion Batteries: A Modified Reference Governor Approach}.
\newblock \bibinfo{journal}{Proc. Am. Control Conf. ACC~2013},
\bibinfo{city}{Washington, DC, USA},
\bibinfo{pages}{5332--5337} (\bibinfo{year}{2013}) 

\bibitem{Perez2014}
\bibinfo{author}{H. Perez}, \bibinfo{author}{N. Shahmohammadhamedani}, \bibinfo{author}{S. Moura},
\newblock \bibinfo{title}{Enhanced Performance of Li-Ion Batteries via Modified Reference Governors and Electrochemical Models}.
\newblock \bibinfo{journal}{IEEE/ASME Trans. Mechatronics},
 \bibinfo{pages}{1--10} (\bibinfo{year}{2014})

\bibitem{Chaturvedi2010}
\bibinfo{author}{N.~A. Chaturvedi}, \bibinfo{author}{R. Klein}, \bibinfo{author}{J. Christensen}, \bibinfo{author}{J. Ahmed}, \bibinfo{author}{A. Kojic},
\newblock \bibinfo{title}{Algorithms for advanced battery-management systems: modeling, estimation, and control challenges for lithium-ion batteries}.
\newblock \bibinfo{journal}{IEEE Control Syst. Mag.}, \bibinfo{volume}{30(3)},
 \bibinfo{pages}{49--68} (\bibinfo{year}{2010})

\bibitem{Moura2014}
\bibinfo{author}{S.~J. Moura}, \bibinfo{author}{H. Perez},
\newblock \bibinfo{title}{Better batteries through electrochemistry}.
\newblock \bibinfo{journal}{ASME Dyn. Syst. and Control Mag.},
 \bibinfo{pages}{15--21} (\bibinfo{year}{2014})

\bibitem{Santhanagopalan2006a}
\bibinfo{author}{S. Santhanagopalan}, \bibinfo{author}{R.~E. White},
\newblock \bibinfo{title}{Online estimation of the state of charge of a lithium ion cell}.
\newblock \bibinfo{journal}{J. Electrochem. Soc.}, \bibinfo{volume}{161(2)}, \bibinfo{pages}{1346--1355}  (\bibinfo{year}{2006})

\bibitem{Smith2008b}
\bibinfo{author}{K. Smith}, \bibinfo{author}{C.~D. Rahn}, \bibinfo{author}{C-Y Wang},
\newblock \bibinfo{title}{Online estimation of the state of charge of a lithium ion cell}.
\newblock \bibinfo{conference}{IEEE International Conference on Control Applications}, \bibinfo{city}{San Antonio, Texas, USA}, \bibinfo{pages}{714--719} (\bibinfo{year}{2008})

\bibitem{DiDomenico2010}
\bibinfo{author}{D. Di~Domenico}, \bibinfo{author}{A. Stefanopoulou}, \bibinfo{author}{G. Fiengo},
\newblock \bibinfo{title}{Lithium-Ion Battery State of Charge and Critical Surface Charge Estimation Using an Electrochemical Model-Based Extended Kalman Filter}.
\newblock \bibinfo{journal}{J. Dyn. Syst., Meas., Control}, \bibinfo{volume}{132},
\bibinfo{pages}{061302} (\bibinfo{year}{2010}) 

\bibitem{Smith2010}
\bibinfo{author}{K. Smith}, \bibinfo{author}{C.~D. Rahn}, \bibinfo{author}{C-Y Wang},
\newblock \bibinfo{title}{Model-Based Electrochemical Estimation and Constraint Management for Pulse Operation of Lithium Ion Batteries}.
\newblock \bibinfo{journal}{IEEE Trans. Control Syst. Technol.}, \bibinfo{volume}{18(3)},
 \bibinfo{pages}{654--663} (\bibinfo{year}{2010})

\bibitem{Moura2013b}
\bibinfo{author}{S.~J. Moura}, \bibinfo{author}{N.~A. Chaturvedi}, \bibinfo{author}{M. Krstic},
\newblock \bibinfo{title}{Adaptive Partial Differential Equation Observer for Battery State-of-Charge/State-of-Health Estimation Via an Electrochemical Model}.
\newblock \bibinfo{journal}{J. Dyn. Syst., Meas., Control}, \bibinfo{volume}{136}, \bibinfo{pages}{011015} (\bibinfo{year}{2013})

\bibitem{Stetzel2015a}
\bibinfo{author}{K.~D. Stetzel}, \bibinfo{author}{L.~L. Aldrich}, \bibinfo{author}{M.~S. Trimboli}, \bibinfo{author}{G.~L. Plett},
\newblock \bibinfo{title}{Electrochemical state and internal variables estimation using a reduced-order physics-based model of a lithium-ion cell and an extended Kalman filter}.
\newblock \bibinfo{journal}{J. Power Sources}, \bibinfo{volume}{278}, \bibinfo{pages}{490--505} (\bibinfo{year}{2015}) 

\bibitem{Zhao2015a}
\bibinfo{author}{S. Zhao}, \bibinfo{author}{A.~M. Bizeray}, \bibinfo{author}{S.~R. Duncan}, \bibinfo{author}{D.~A. Howey},
\newblock \bibinfo{title}{Performance evaluation of an extended Kalman filter for state estimation of a pseudo-2D thermal-electrochemical lithium-ion battery model}.
\newblock \bibinfo{journal}{ASME Dyn. Syst. Control Conf., Proc.}, \bibinfo{city}{Columbus, Ohio, USA}, \bibinfo{pages}{1--5} (\bibinfo{year}{2015}) 

\bibitem{Northrop2011b}
\bibinfo{author}{P.~W.~C. Northrop}, \bibinfo{author}{V. Ramadesigan}, \bibinfo{author}{S. De}, \bibinfo{author}{V.~R. Subramanian},
\newblock \bibinfo{title}{Coordinate Transformation, Orthogonal Collocation, Model Reformulation and Simulation of Electrochemical-Thermal Behavior of Lithium-Ion Battery Stacks}.
\newblock \bibinfo{journal}{J. Electrochem. Soc.}, \bibinfo{volume}{158(12)}, \bibinfo{pages}{A1461--A1477} (\bibinfo{year}{2011})

\bibitem{Schmidt2010}
\bibinfo{author}{A.~P. Schmidt}, \bibinfo{author}{M. Bitzer}, \bibinfo{author}{A.~W. Imre}, \bibinfo{author}{L. Guzzella},
\newblock \bibinfo{title}{Experiment-driven electrochemical modeling and systematic parameterization for a lithium-ion battery cell}.
\newblock \bibinfo{journal}{J. Power Sources}, \bibinfo{volume}{195(15)}, \bibinfo{pages}{5071-5080} (\bibinfo{year}{2010})

\bibitem{Ramadesigan2011}
\bibinfo{author}{V. Ramadesigan}, \bibinfo{author}{K. Chen}, \bibinfo{author}{N.~A. Burns}, \bibinfo{author}{V. Boovaragavan}, \bibinfo{author}{R.~D. Braatz}, \bibinfo{author}{V.~R. Subramanian},
\newblock \bibinfo{title}{Parameter Estimation and Capacity Fade Analysis of Lithium-Ion Batteries Using Reformulated Models}.
\newblock \bibinfo{journal}{J. Electrochem. Soc.}, \bibinfo{volume}{158(9)}, \bibinfo{pages}{A1048-A1054} (\bibinfo{year}{2011})

\bibitem{Forman2011c}
\bibinfo{author}{J.~C. Forman}, \bibinfo{author}{S.~J. Moura}, \bibinfo{author}{J.~L. Stein}, \bibinfo{author}{H.~K. Fathy},
\newblock \bibinfo{title}{Genetic parameter identification of the Doyle-Fuller-Newman model from experimental cycling of a LiFePO\textsubscript{4} battery}.
\newblock \bibinfo{conference}{Proc. Am. Control Conf. ACC~2011}, \bibinfo{city}{San Francisco, CA, USA}, \bibinfo{pages}{362--369} (\bibinfo{year}{2011})

\bibitem{Forman2012}
\bibinfo{author}{J.~C. Forman}, \bibinfo{author}{S.~J. Moura}, \bibinfo{author}{J.~L. Stein}, \bibinfo{author}{H.~K. Fathy},
\newblock \bibinfo{title}{Genetic identification and fisher identifiability analysis of the Doyle-Fuller-Newman model from experimental cycling of a LiFePO\textsubscript{4} cell}.
\newblock \bibinfo{journal}{J.Power Sources}, \bibinfo{volume}{210}, \bibinfo{pages}{263--275} (\bibinfo{year}{2012})

\bibitem{Marcicki2013a}
\bibinfo{author}{J. Marcicki}, \bibinfo{author}{M. Canova}, \bibinfo{author}{A.~T. Conlisk}, \bibinfo{author}{G. Rizzoni},
\newblock \bibinfo{title}{Design and parametrization analysis of a reduced-order electrochemical model of graphite/LiFePO\textsubscript{4} cells for SOC/SOH estimation}.
\newblock \bibinfo{journal}{J. Power Sources}, \bibinfo{volume}{237}, \bibinfo{pages}{310--324} (\bibinfo{year}{2013}) 

\bibitem{Jiang2011}
\bibinfo{author}{S. Jiang},
\newblock \bibinfo{title}{A Parameter Identification Method for a Battery Equivalent Circuit Model}.
\newblock \bibinfo{journal}{Soc. Automot. Eng. Tech. Pap. Ser.},
\bibinfo{pages}{1--9} (\bibinfo{year}{2011})
 
\bibitem{Nazer2012}
\bibinfo{author}{R.~A. Nazer}, \bibinfo{author}{V. Cattin}, \bibinfo{author}{P. Granjon}, \bibinfo{author}{M. Montaru},
\newblock \bibinfo{title}{A new optimization algorithm for a li-ion battery equivalent electrical equivalent circuit identification}.
\newblock \bibinfo{conference}{9\textsuperscript{th} International Conference of Modeling, Optimization and Simulation - MOSIM’12}, \bibinfo{city}{Bordeaux, France}, \bibinfo{pages}{1--7} (\bibinfo{year}{2012})

\bibitem{Jang2011}
\bibinfo{author}{J. Jang}, \bibinfo{author}{J. Yoo},
\newblock \bibinfo{title}{Equivalent circuit evaluation method of lithium polymer battery using bode plot and numerical analysis}.
\newblock \bibinfo{journal}{IEEE Trans. Energy Convers.}, \bibinfo{volume}{26(1)},
 \bibinfo{pages}{290--298} (\bibinfo{year}{2011})

\bibitem{Moubayed2008}
\bibinfo{author}{N. Moubayed}, \bibinfo{author}{J. Kouta}, \bibinfo{author}{A. El-Ali}, \bibinfo{author}{H. Dernayka},\bibinfo{author}{R. Outbib},
\newblock \bibinfo{title}{Parameter identification of the lead-acid battery model}.
\newblock \bibinfo{conference}{IEEE Photovoltaic Spec. Conf., 33\textsuperscript{rd}},
 \bibinfo{pages}{1--6} (\bibinfo{year}{2008})

\bibitem{Doyle1993}
\bibinfo{author}{M. Doyle}, \bibinfo{author}{T.~F. Fuller}, \bibinfo{author}{J. Newman},
\newblock \bibinfo{title}{Modeling of Galvanostatic Charge and Discharge of the Lithium/Polymer/Insertion Cell}.
\newblock \bibinfo{journal}{J. Electrochem. Soc.}, \bibinfo{volume}{140(6)}, \bibinfo{pages}{1526-1533} (\bibinfo{year}{1993})

\bibitem{Atlung1979}
\bibinfo{author}{S. Atlung}, \bibinfo{author}{K. West}, \bibinfo{author}{T. Jacobsen},
\newblock \bibinfo{title}{Dynamic aspects of solid solution cathodes for electrochemical power sources}.
\newblock \bibinfo{journal}{J. Electrochem. Soc.}, \bibinfo{volume}{126(8)}, \bibinfo{pages}{1311--1321} (\bibinfo{year}{1979})

\bibitem{Ning2004}
\bibinfo{author}{G. Ning}, \bibinfo{author}{B.~N. Popov},
\newblock \bibinfo{title}{Cycle Life Modeling of Lithium-Ion Batteries}.
\newblock \bibinfo{journal}{J. Electrochem. Soc.}, \bibinfo{volume}{151(10)}
 \bibinfo{pages}{A1584-A1591}, (\bibinfo{year}{2004})
 
\bibitem{Barsoukov2005}
\bibinfo{author}{E. Barsoukov}, \bibinfo{author}{J.~R. Macdonald},
\newblock \bibinfo{title}{Impedance Spectroscopy: Theory, Experiment, and Applications}.
\newblock \bibinfo{publisher}{John Wiley and Sons, Inc.}, \bibinfo{city}{Hoboken, NJ, USA}, \bibinfo{pages}{1--595} (\bibinfo{year}{2005})

\bibitem{Tippman2014}
\bibinfo{author}{S. Tippmann}, \bibinfo{author}{D. Walper}, \bibinfo{author}{L. Balboa}, \bibinfo{author}{B. Spier},\bibinfo{author}{W.~G. Bessler},
\newblock \bibinfo{title}{Low-temperature charging of lithium-ion cells part I: Electrochemical
modeling and experimental investigation of degradation behavior}.
\newblock \bibinfo{journal}{J. Power Sources}, \bibinfo{volume}{252},
 \bibinfo{pages}{305--316} (\bibinfo{year}{2014})
 
 \bibitem{Murbach2017}
\bibinfo{author}{M.~D. Murbach}, \bibinfo{author}{V.~W. Hu}, \bibinfo{author}{D.~T. Schwartz},
\newblock \bibinfo{title}{The Impedance Analyzer: An Open-Source, Web-Based Tool for Sophisticated Interrogation of Experimental EIS Spectra}.
\newblock \bibinfo{conference}{ECS Meeting Abstracts},
 \bibinfo{pages}{320--320} (\bibinfo{year}{2017})

\bibitem{Santhanagopalan2007}
 \bibinfo{author}{S. Santhanagopalan}, \bibinfo{author}{Q. Guo}, \bibinfo{author}{R.~E. White},
 \newblock \bibinfo{title}{Parameter Estimation and Model Discrimination for a Lithium-Ion Cell}.
 \newblock \bibinfo{journal}{J. Electrochem. Soc.}, \bibinfo{volume}{154(3)}
 (\bibinfo{year}{2007}) \bibinfo{pages}{A198--A206}

\bibitem{Bellman1970}
\bibinfo{author}{R. Bellman}, \bibinfo{author}{K.~J. Astrom},
\newblock \bibinfo{title}{On structural identifiability}.
\newblock \bibinfo{journal}{Math. Biosci.}, \bibinfo{volume}{7},
\bibinfo{pages}{329--339} (\bibinfo{year}{1970})

\bibitem{Ljung1999}
\bibinfo{author}{L. Ljung}
\newblock \bibinfo{title}{System Identification: Theory for the User}.
\newblock \bibinfo{publisher}{Prentice Hall PTR} (\bibinfo{year}{1999})

\bibitem{Alavi2016}
\bibinfo{author}{S.~S.~M Alavi}, \bibinfo{author}{A. Mahdi}, \bibinfo{author}{, S.~J. Payne}, \bibinfo{author}{D.~A. Howey},
\newblock \bibinfo{title}{Identifiability of Generalized Randles Circuit Models}.
\newblock \bibinfo{journal}{ArXiv e-prints}, \bibinfo{url}{arXiv:1505.00153 [math.OC]} (\bibinfo{year}{2016})


\bibitem{Birkl2016}
\bibinfo{author}{C.~R. Birkl}, \bibinfo{author}{E. McTurk}, \bibinfo{author}{M.~R. Roberts}, \bibinfo{author}{P.~G. Bruce}, \bibinfo{author}{D.~A. Howey},
\newblock \bibinfo{title}{A Parametric Open Circuit Voltage Model for Lithium Ion Batteries}.
\newblock \bibinfo{journal}{J. Electrochem. Soc.}, \bibinfo{volume}{162(12)}, \bibinfo{pages}{A2271--A2280} (\bibinfo{year}{2015})

\bibitem{McTurk2015}
\bibinfo{author}{E. McTurk}, \bibinfo{author}{C.~R. Birkl},  \bibinfo{author}{M.~R. Roberts}, \bibinfo{author}{D.~A. Howey}, \bibinfo{author}{P.~G. Bruce},
\newblock \bibinfo{title}{Minimally Invasive Insertion of Reference Electrodes into Commercial Lithium-Ion Pouch Cells}.
\newblock \bibinfo{journal}{ECS Electrochem. Lett.}, \bibinfo{volume}{4(12)}, \bibinfo{pages}{A145--A147} (\bibinfo{year}{2015})


\bibitem{Haran1998}
\bibinfo{author}{B.~S. Haran}, \bibinfo{author}{B.~N. Popov}, \bibinfo{author}{R.~E. White},
\newblock \bibinfo{title}{Determination of the hydrogen diffusion coefficient in metal hydrides by impedance spectroscopy}.
\newblock \bibinfo{journal}{J. Power Sources}, \bibinfo{volume}{75}
(\bibinfo{year}{1998}) \bibinfo{pages}{56--63}

\bibitem{Guo2002}
\bibinfo{author}{Q. Guo}, \bibinfo{author}{V.~R. Subramanian}, \bibinfo{author}{J.~W. Weidner}, \bibinfo{author}{R.~E. White},
\newblock \bibinfo{title}{Estimation of Diffusion Coefficient of Lithium in Carbon Using AC Impedance Technique}.
\newblock \bibinfo{journal}{J. Electrochem. Soc.}, \bibinfo{volume}{149(3)}
(\bibinfo{year}{2002}) \bibinfo{pages}{A307--A318}

\bibitem{Sikha2007}
\bibinfo{author}{G. Sikha}, \bibinfo{author}{R.~E. White},
\newblock \bibinfo{title}{Analytical Expression For the Impedance Response of an Insertion Electrode Cell}.
\newblock \bibinfo{journal}{J. Electrochem. Soc.}, \bibinfo{volume}{154(1)}
(\bibinfo{year}{2007}) \bibinfo{pages}{A43--A54}



\bibitem{Buller2003a}
\bibinfo{author}{S. Buller}, \bibinfo{author}{M. Thele}, \bibinfo{author}{E. Karden}, \bibinfo{author}{R.W. De Doncker},
\newblock \bibinfo{title}{Impedance-based non-linear dynamic battery modeling for automotive applications}.
\newblock \bibinfo{journal}{J. Power Sources}, \bibinfo{volume}{113(2)}, \bibinfo{pages}{422-430} (\bibinfo{year}{2003})

\bibitem{Howey2014}
\bibinfo{author}{D.~A. Howey}, \bibinfo{author}{P.~D. Mitcheson}, \bibinfo{author}{V. Yufit}, \bibinfo{author}{G.~J. Offer}, \bibinfo{author}{N.~P. Brandon},
\newblock \bibinfo{title}{Online measurement of battery impedance using motor controller excitation}.
\newblock \bibinfo{journal}{IEEE Trans. Veh. Technol.}, \bibinfo{volume}{63(6)}, \bibinfo{pages}{2557--2566} (\bibinfo{year}{2014})

\bibitem{Ecker2015}
\bibinfo{author}{M. Ecker}, \bibinfo{author}{T.~K.~D. Tran}, \bibinfo{author}{P. Dechent}, \bibinfo{author}{S. Kabitz},\bibinfo{author}{A. Warnecke},\bibinfo{author}{D.~U. Sauer},
\newblock \bibinfo{title}{Parameterization of a Physico-Chemical Model of a Lithium-Ion Battery: I. Determination of Parameters}.
\newblock \bibinfo{journal}{J. Electrochem. Soc.}, \bibinfo{volume}{162(9)}, \bibinfo{pages}{A1836--A1848} (\bibinfo{year}{2015})

\bibitem{Fuller1994}
\bibinfo{author}{T.~F. Fuller}, \bibinfo{author}{M. Doyle}, \bibinfo{author}{J. Newman},
\newblock \bibinfo{title}{Simulation and Optimization of the Dual Lithium Ion Insertion Cell}.
\newblock \bibinfo{journal}{J. Electrochem. Soc.}, \bibinfo{volume}{141(1)}, \bibinfo{pages}{1--10} (\bibinfo{year}{1994}) 

\bibitem{Guo2011}
\bibinfo{author}{M. Guo}, \bibinfo{author}{G. Sikha}, \bibinfo{author}{R.~E. White},
\newblock \bibinfo{title}{Single-Particle Model for a Lithium-Ion Cell: Thermal Behavior}.
\newblock \bibinfo{journal}{J. Electrochem. Soc.}, \bibinfo{volume}{158(2)}, \bibinfo{pages}{A122--A132} (\bibinfo{year}{2011})

\bibitem{Zhao2017}
\bibinfo{author}{S. Zhao}, \bibinfo{author}{S.~R. Duncan}, \bibinfo{author}{D.~A. Howey},
\newblock \bibinfo{title}{Observability Analysis and State Estimation of Lithium-Ion Batteries in the Presence of Sensor Biases}.
\newblock \bibinfo{journal}{IEEE Trans. Control Syst. Technol.}, \bibinfo{volume}{25(1)}, \bibinfo{pages}{326--333} (\bibinfo{year}{2017})

\bibitem{Bizeray2016}
\bibinfo{author}{A.~M. Bizeray}, \bibinfo{author}{J. Reniers}, \bibinfo{author}{D.~A. Howey},
\newblock \bibinfo{title}{Spectral li-ion SPM}. (\bibinfo{year}{2016}), 
\bibinfo{url}{http://doi.org/10.5281/zenodo.212178}

% %% To cite about EIS


% \bibitem[]{}
% \bibinfo{author}{}, \bibinfo{author}{}, \bibinfo{author}{}, \bibinfo{author}{},
% \newblock \bibinfo{title}{}.
% \newblock \bibinfo{journal}{}, \bibinfo{volume}{}
% (\bibinfo{year}{}) \bibinfo{pages}{}

\end{thebibliography}
\end{document}